\documentclass[%
 reprint,
 amsmath,amssymb,
aps,
prl,
floatfix
]{revtex4-2}
\usepackage[utf8]{inputenc}
\usepackage{graphicx}
\usepackage{xcolor}
\usepackage[english]{babel}
\usepackage{mathrsfs}
\usepackage{mathtools}
\usepackage{amsfonts}
\usepackage{booktabs}
\usepackage{amsthm}
\usepackage{dsfont}
\usepackage{setspace}
\usepackage{hyperref}

\begin{document}
\title{A Toolbox for Quantifying Memory in Dynamics
  Along Reaction Coordinates}

\author{Alessio Lapolla}
\author{Alja\v{z} Godec}%
 \email{agodec@mpibpc.mpg.de}
\affiliation{%
  Mathematical bioPhysics Group, Max Planck Institute for
  Biophysical Chemistry, G\"{o}ttingen 37077, Germany
}%

\date{\today}

\begin{abstract}
Memory effects in time-series of experimental observables
are ubiquitous, have important consequences for the
interpretation of kinetic data, and may even affect the function of
biomolecular nanomachines such as enzymes. 
Here we propose a set of complementary methods for quantifying
conclusively the magnitude and duration of memory in a time series of
a reaction coordinate. The toolbox is general, robust, easy to
use, and does not rely on any underlying microscopic model. As a proof
of concept we apply it to the analysis of memory in the
dynamics of the end-to-end distance of the analytically solvable
Rouse-polymer model, an experimental time-series of
extensions of a single DNA hairpin measured \textcolor{black}{by} optical tweezers,
\textcolor{black}{ and
  the fraction of native contacts in a small
  protein probed by atomistic Molecular Dynamics simulations.}
\end{abstract}

\maketitle

The dynamics of complex, high-dimensional physical systems such as
complex biomolecules is frequently described by means of memory-less, Markovian diffusion
along a low-dimensional reaction
coordinate
\cite{Best,Portman,Baron,Faradjian,Berezhkovskii_2005,Hummer_2005,Best_2009,Zhang_2016,Dima_2017,Berne_1988,Hummer}. Such
simplified models often accurately describe selected observations in
experiments \cite{Dudko_2008,neupane_protein_2016,neupane_direct_2016,Gladrow,Thorneywork}
and computer simulations \cite{Best_2009,Hummer_2005,Best}. However,
as soon as latent, hidden degrees of freedom that become projected out
do not relax instantaneously on the time scale we observe the
reaction coordinate \cite{lapolla_manifestations_2019}, or the reaction
coordinate does not locally equilibrate in 
meta-stable meso-states \cite{Hartich_2020},
almost any projection of high-dimensional dynamics onto a lower
dimensional coordinate introduces memory \cite{van_kampen_remarks_1998,Wolynes,
  zwanzig_nonequilibrium_2010,Dima_2013,lapolla_manifestations_2019,Dima_2019,Meyer_2020,Sollich_2020,Krueger_2020,Hartich_2020}.     

Memory effects can have intriguing manifestations in the evolution of both,
ensemble-
\cite{Xie_2005,lapolla_manifestations_2019,Lapolla_2020,Lapolla_2021,Lapolla_JCP}
and time-averaged observables
\cite{Lapolla_2018,lapolla_manifestations_2019}, and are often  particularly well-pronounced in
observations that reflect, or couple to, intra-molecular distances in
conformationally flexible biomolecules \cite{Wolynes,Xie_2004,Xie_2005,JCS_2008,Dima_2013,Dill_2014,Hu_2015,Dima_2019,Meyer_2020,Hartich_2020,sangha_proteins_2009,
  avdoshenko_theoretical_2017, cote_anomalous_2012,
  grossman-haham_slow_2018, pyo_memory_2019}. Moreover, if the
dynamics is ergodic in the sense that the system relaxes to a unique equilibrium probability
density function from any initial condition  (i.e.\ the reaction
coordinate has a unique free energy landscape) then the memory is
necessarily transient \cite{lapolla_manifestations_2019}. Whether
or not memory is in fact relevant depends on how its extent compares to the
relaxation time and whether or not the latter is reached in an
experiment.    
If the extent of memory is comparable to, or longer
than, the time-scale on which biomolecules operate, such
e.g.\ enzymes catalyzing chemical reactions \cite{enzyme,enzyme_2}, 
non-Markovian effects shape biological function. 

It is therefore important to assess the presence and duration of memory
effects in the dynamics along reaction coordinates. An elegant ``test
of Markovianity'' of a reaction coordinate has
recently been proposed by Berezhkovskii and
Makarov, who considered the behavior of transition paths
\cite{berezhkovskii_single-molecule_2018}. The authors provide a pair
of inequalities whose violation conclusively reflects that the
dynamics is non-Markovian. However, memory-effects are typically
transient \cite{lapolla_manifestations_2019} although their extent may exceed the duration of experimental
observations \cite{Hu_2015}. There is thus a need to determine not
only the presence of memory in a time-series of a reaction coordinate but also its extent and attenuation on
different time-scales.

Here, we fill this gap by providing a toolbox for
quantifying \textcolor{black}{the magnitude and duration of} memory in a time-series of a reaction
coordinate. We propose a set of model-free complementary methods that
are easy to use and suited to treat reaction coordinates with
arbitrary dimensionality. As a proof of concept we apply these methods
to the analysis of an experimental time-series of the extension of a
DNA-hairpin measured 
by optical tweezers,
\textcolor{black}{the fraction of native
  contacts in a protein probed by atomistic
  Molecular Dynamics (MD) simulations,}  and the exactly-solvable Rouse
model of polymer chain.

\emph{Theory.---} \textcolor{black}{Our approach is twofold -- (i) we
  quantify violations of the Chapman-Kolmogorov equation 
 in a time series of the monitored \emph{true dynamics}, 
  and (ii) compare the true dynamics to a constructed \emph{nominally 
  memory-less diffusion} in the free energy-
and diffusion-landscape of the true dynamics. This assumes all hidden
degrees of freedom to be at equilibrium constrained by the
instantaneous value of the observable}. 


Let $q_t$ with $0\le t\le T$ denote the \textcolor{black}{monitored} time-series
of the reaction coordinate and $q^{\rm M}_t$ the
\textcolor{black}{constructed} Markovian
series. Without any loss of generality we assume that the reaction
coordinate is one-dimensional -- the generalization to multiple
dimensions is straightforward. We assume $q_t$ and $q^{\rm M}_t$ to be ergodic with
an equilibrium probability density $p_{\rm eq}(q)$ that is by
construction identical for both processes. Let $G(q,t|q_0)=\langle \delta(q_t-q)\rangle_{q_0}$ denote the
probability density that the reaction coordinate evolving from
$q_{t=0}=q_0$ is found at time $t$ to have a value in an infinitesimal
neighborhood of $q$ and $G^{\rm M}(q,t|q_0)=\langle \delta(q^{\rm
  M}_t-q)\rangle_{q_0}$ the 
Markovian
counterpart, where $\delta(x)$ denotes Dirac's delta function
and the angular brackets $\langle \cdot\rangle_{q_0}$ the average over
all realizations of $q_t$ evolving from $q_0$. We then have
$\lim_{t\to\infty}G(q,t|q_0)=\lim_{t\to\infty}G_M(q,t|q_0)=p_{\rm
  eq}(q)$ as a result of ergodicity. In practice the limits are
achieved as soon as $t$ becomes sufficiently larger than the
relaxation time $t_{\rm rel}$, i.e. $t\gtrsim t_{\rm rel}$, which may
or may not be reached in an experiment. Note that the relaxation times
of the true and Markovian reference process are typically different \cite{lapolla_manifestations_2019,Lapolla_JCP}. 

We use two descriptors. The first is the
Kullback-Leibler divergence between the transition probabilities of
the true and a reference process defined as \cite{s._kullback_information_1951}
\begin{equation}
  \mathcal{D}^{\rm a}_{q_0}(t)\equiv\int dq
  G(q,t|q_0)\ln[G(q,t|q_0)/G^{\rm a}(q,t|q_0)],
  \label{klddiv}
\end{equation}
where $a={\rm CK, M}$ denotes the particular kind of reference process
that we detail below. By construction $\mathcal{D}^{\rm
  a}_{q_0}(t)\ne 0$ if and only if $G(q,t|q_0)\ne G^{\rm a}(q,t|q_0)$
and thus non-zero values of $\mathcal{D}^{\rm
  a}_{q_0}(t)$ reflect memory in the dynamics of $q_t$.

When $q_t$ reaches equilibrium in the course of
the experiment we also consider the normalized equilibrium autocorrelation function defied as
\begin{equation}
C_{,{\rm M}}(t)\equiv \frac{\langle q_tq_0\rangle_{,{\rm M}}-\langle q\rangle_{\rm eq}^2}{\langle q^2\rangle_{\rm eq}-\langle q\rangle_{\rm eq}^2},
  \label{autocorr}
\end{equation}
where we have introduced
\begin{eqnarray}
\langle q_tq_0\rangle_{,{\rm M}}&\equiv&\int\int qq_0G^{,{\rm M}}(q,t|q_0)p_{\rm
  eq}(q_0)dqdq_0\nonumber\\
&\overset{T\gg t_{\rm rel}}{=}&(T-t)^{-1}\int_0^{T-t}
q_{\tau+t}q_\tau d\tau,\,\,t\ll T, \nonumber\\
\langle q^n\rangle_{\rm eq} &\equiv&\int q^np_{\rm
  eq}(q)dq,\,\,\, {\rm for}\,\,\, n=1,2\nonumber\\
&\overset{T\gg t_{\rm rel}}{=}&T^{-1}\int_0^Tq^n_\tau d\tau
  \label{exp}
\end{eqnarray}
where the 
definitions in terms of time-averages hold
when trajectories are much longer than the relaxation time,
i.e. $T\gg t_{\rm rel}$. The absence of
an index refers to the true process and 
${\rm M}$ to the \textcolor{black}{constructed}
Markovian counterpart.

We consider two distinct reference processes. The first
one \textcolor{black}{is a mathematical construction based on the
Chapman-Kolmogorov equation (i.e. $a={\rm CK}$)} that we may write as 
\begin{equation}
G_{\tau}^{\rm CK}(q,t|q_0)\equiv \int G(q,t-\tau|q')G(q',\tau|q_0)dq',  
\label{CKol}
\end{equation}
because the Green's function of a time-homogeneous Markov process is time-translation
invariant, $G(q,t-\tau|q')=G(q,t|q',\tau)$, \textcolor{black}{and $G_{\tau}^{\rm
  CK}(q,t|q_0)=G(q,t|q_0)$ independent of $\tau$}  \cite{gardiner_c.w._handbook_1985}. The physical interpretation of
\textcolor{black}{Eq.~\eqref{CKol}, which is exact for Markov processes,} is that we observe the true dynamics $q_t$ until time
$\tau$ and then instantaneously reset the memory (if any) to zero.

If $q_t$ is indeed memoryless we have 
$G_{\tau}^{\rm
  CK}(q,t|q_0)=G(q,t|q_0)$ for any $\tau$ and thus
$\mathcal{D}^{\rm CK}_{\tau,q_0}(t)=0$ for any $t$ and $\tau$. If $G_{\tau}^{\rm
  CK}(q,t|q_0)\ne G(q,t|q_0)$ for some $t$ and $\tau$ then
$q_t$ is conclusively non-Markovian and $\mathcal{D}^{\rm CK}_{\tau,q_0}(t)>0$, but the converse is not
true. Namely, there exist non-Markovian processes that satisfy the
Chapman-Kolmogorov equation
\cite{lapolla_manifestations_2019,feller_non-markovian_1959}. Note
that this method does not require $q_t$ to reach equilibrium during
an experiment and requires only $G(q,t|q_0)$ that is straightforward to
determine from a time series $q_t$ \textcolor{black}{given sufficient
  data}. 
If equilibrium is reached,
$\mathcal{D}^{\rm CK}_{\tau,q_0}(t\gg t_{\rm
  rel})\simeq 0$ for any $q_0$.
%
%
By analyzing
$\mathcal{D}^{\rm CK}_{\tau,q_0}(t)$ we can quantify the
degree and range of memory 
as a function of $\tau$ and $q_0$ which
we demonstrate below.

In the second method we construct from $q_t$ a 
Markovian
time-series $q^{\rm M}_t$ \textcolor{black}{(i.e. $a={\rm M}$)} evolving under the influence of the potential of mean
force $w(q)\equiv -k_{\rm B}T\ln p_{\rm eq}(q)$ according to the
\textcolor{black}{thermodynamically consistent anti-It\^o (i.e.\ 
  post-point) \cite{Hartich_2020}}
Langevin equation
\begin{equation}
  \frac{d}{dt}q_t^{\rm M}=\textcolor{black}{D(q_t^{\rm M})}f(q_t^{\rm
    M})/k_{\rm B}T+\textcolor{black}{\sqrt{2D(q_t^{\rm M})}}\circledast \xi_t,
  \label{Langevin}
  \end{equation}
where $f(q_t^{\rm M})\equiv-k_{\rm B}T \partial_{q}\ln p_{\rm eq}(q)|_{q=q_t^{\rm M}}$ and $\xi_t$ denotes zero mean Gaussian white noise with covariance
$\langle \xi_t\xi_{t'}\rangle=\delta(t-t')$,
$\textcolor{black}{D(q_t)}$ is the diffusion
\textcolor{black}{landscape and $\circledast$ is the anti-It\^o or
  Klimontovich product \cite{Kli} (see Supplementary
  Material (SM) \footnote{See Supplemental Material at [...] for
a discretization of the
anti-It\^o Langevin equation \eqref{Langevin}, exact results for the
Rouse polymer, details about MD simulations,
the fraction of native contacts, the estimation of the diffusion
landscape $D(q)$, and a
description of the uncertainty quantification.} for the discretized version
  of Eq.~\eqref{Langevin}).}
This method assumes the ability to determine the
equilibrium probability density $p_{\rm eq}(q)$ and thus requires $q_t$ to reach
equilibrium. In the simplest model 
the diffusion coefficient does not depend on $q$ \textcolor{black}{and we may interpret
Eq.~\eqref{Langevin} according to It\^o}.  
However, \textcolor{black}{this may not be the case (see below), and} 
we note that the best possible Markovian approximation 
\textcolor{black}{includes a positional dependence} 
\cite{Szabo}. Efficient methods have been developed
\textcolor{black}{to infer $D(q)$}
\cite{Dima_2017,Berne_1988,Hummer}.

 On the level of the probability density function Eq.~(\ref{Langevin})
 corresponds to the Fokker-Planck equation 
\begin{equation}
  \partial_t G^{\rm M}(q,t|q_0)=\partial_q D(q)[\partial_q-
f(q)/k_{\rm B}T]G^{\rm M}(q,t|q_0),
  \label{FPE}
  \end{equation}
with initial condition $G^{\rm M}(q,0|q_0)=\delta(q-q_0)$ and natural
boundary conditions imposed by the underlying physics. Depending on
the specific problem $G^{\rm M}(q,t|q_0)$ can be found by a numerical
integration of the Langevin equation and subsequent histogram
analysis, i.e. $G^{\rm M}(q,t|q_0)=\langle \delta(q^{\rm M}_t-q)\rangle_{q_0}$, or
by \textcolor{black}{projecting the full dynamics or directly solving
  Eq.~(\ref{FPE}) as done
  e.g.\ for
  polymers
 \cite{sunagawa_theory_1975}, single-file models
\cite{lapolla_manifestations_2019,Lapolla_2020}, and in the 
literature on
persistence \cite{Satya1,Satya2}
in diffusive and critical dynamics
\cite{Bray_1994,Derrida,Satya_c1,Satya_critical,Satya1,Satya2}}. Below we illustrate both
approaches. 

\emph{End-to-end distance of a Rouse polymer.---}
As a first example we consider a
Rouse polymer chain with $N+1$ beads ($N$ bonds) in absence of hydrodynamic
interactions  \cite{rouse_theory_1953, ahn_bead-spring_1993}
 and focus on the end-to-end distance as the reaction
coordinate, i.e. $q_t\equiv|\mathbf{r}_1-\mathbf{r}_{N+1}|$, \textcolor{black}{which is
known to be non-Markovian. 
The model is exactly solvable and the explicit results for $C(t),
C_M(t),G(q,t|q_0),G^{\rm M}(q,t|q_0)$ and $G_{\tau}^{\rm CK}(q,t|q_0)$
are all given in \cite{Note1}}. We express time in
units of $t_{\rm Kuhn}$, the characteristic diffusion time of a
Kuhn-segment, i.e. $t_{\rm Kuhn}=b^2/D$, where $b$ is the Kuhn-length
and $D$ the diffusion coefficient of a bead.  

\begin{figure}
    \centering
    \includegraphics[width=8.6cm]{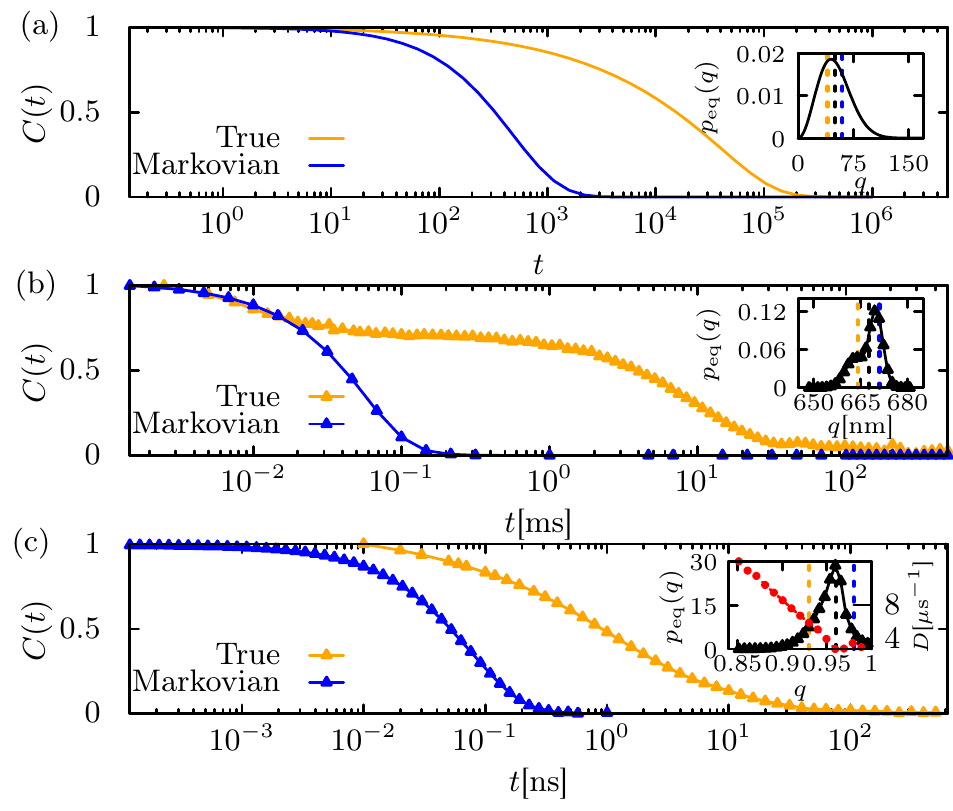}
    \caption{\textcolor{black}{Autocorrelation function of the true
      dynamics (orange)
      and its Markovian approximation (blue) for (a) a Rouse-polymer
      with 1000 monomers with time expressed in units of the diffusion
      time of a Kuhn-segment $t_{\rm Kuhn}$; (b) the extension of a
      DNA-hairpin, and (c) the fraction of native contacts in the WW-domain
of protein 2F21. The black line in the inset depicts the respective
        equilibrium probability density function $p_{\rm eq}(q)$ and
        the red one $D(q)$. The
        dashed lines depict the initial conditions we consider in Fig.~\ref{fig:klchapkol}.}} 
    \label{fig:autocorrelation}
\end{figure}

A comparison of the autocorrelation function of the true dynamics
and \textcolor{black}{its} Markovian approximation $C_M(t)$
is shown in
Fig.~\ref{fig:autocorrelation}a, with the inset depicting the corresponding
equilibrium probability density $p_{\rm eq}(q)$.
Note that when the free energy landscape $w(q)$ overestimates the confining
effect of hidden degrees of freedom on $q_t$ the Markovian approximation
overestimates the relaxation rate
(e.g.\ \cite{lapolla_manifestations_2019}; see also \cite{Note1}). Namely, the Markovian approximation assumes the hidden degrees
of freedom to remain at equilibrium at all times, whereas the actual
instantaneous, fluctuating
restoring force on $q_t$ is in this case smaller than the force
arising from $w(q)$.

\begin{figure*}
    \centering
    \includegraphics[width=16.cm]{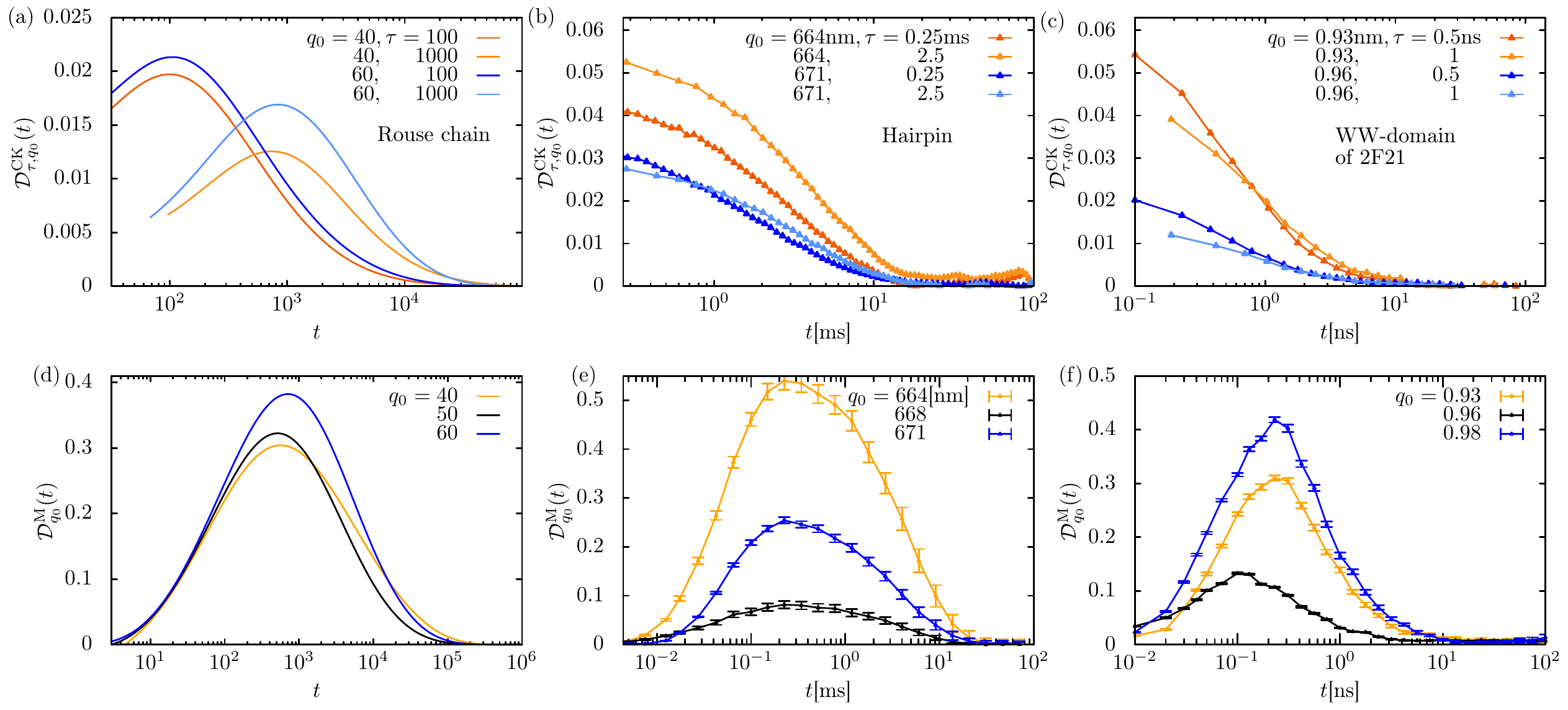}
    \caption{\textcolor{black}{Kullback-Leibler divergence $\mathcal{D}^{\rm a}_t$ in Eq.~(\ref{klddiv})
        between true Green's function $G(q,t|q_0)$ and (a-c) the Chapman-Kolmogorov
      Eq.~(\ref{CKol}) (i.e. $a={\rm CK}$), and (d-f)
      the Markovian approximation $G^{\rm M}(q,t|q_0)$ corresponding to
      the Langevin Eq.~ \eqref{Langevin} (i.e. $a={\rm M}$)
      as a function of time
      $t$ for (a) and (d) the
      Rouse-polymer with 1000 beads evolving from several initial
      conditions $q_0$,
      (b) and (e) the extension of a
      DNA-hairpin evolving from several initial
      conditions
      within a bin of thickness
      $1$nm centered at $q_0$,
      and (c) and (f) the fraction of native
      contacts in the WW-domain
of protein 2F21 for several $q_0$; the error bars depict the standard
      deviation obtained by systematically neglecting $\sim$20\%\ (in case
      of the hairpin) and $\sim$40\%\ (in case
      of the protein) of the data.
    Due to
      the particular construction of Eq.~(\ref{CKol}) times shorter
      than depicted are not accessible due to numerical instability
      or poor statistics.}}
    \label{fig:klchapkol}
\end{figure*}

The Chapman-Kolmogorov-construct for the Rouse polymer, $G_{\tau}^{\rm
  CK}(q,t|q_0)$,
\textcolor{black}{(given explicitly in the \cite{Note1})}
differs from the true $G(q,t|q_0)$
for all expect
large values of $t-\tau$. A quantification of the discrepancy between the true and
``Chapman-Kolmogorov'' evolution of the end-to-end distance of the
Rouse-polymer in terms of the Kullback-Leibler divergence (\ref{klddiv}) is shown in Fig.~\ref{fig:klchapkol}a.
A typical time evolution of $\mathcal{D}^{\rm CK}_{\tau,q_0}(t)$
gradually increases from zero, reaches a maximum and afterwards
returns back to $0$, which reflects the gradual build-up and
attenuation of memory because $q_t$ ``remembers'' the initial
condition of the hidden degrees of freedom
\cite{lapolla_manifestations_2019}. As a result, the
Chapman-Kolmogorov Green's function  $G_{\tau}^{\rm CK}(q,t|q_0)$
fails to predict the true evolution of $q_t$, and $\mathcal{D}^{\rm CK}_{\tau,q_0}(t)$
constructed this way depends on both, $\tau$ and initial condition
$q_0$.
For the Rouse-polymer with 1000 beads $\mathcal{D}^{\rm
  CK}_{\tau,q_0}(t)\ne 0$ at least up to
$t\sim 10^4\times t_{\rm Kuhn}$.

Next we examine $\mathcal{D}^{\rm M}_{q_0}(t)$, the Kullback-Leibler divergence (\ref{klddiv}) between
the true Green's function $G(q,t|q_0)$ and the Markovian approximation
corresponding to the white-noise Markovian diffusion in the exact free
energy landscape (i.e. Eq.~(\ref{Langevin})). The results are shown in
Fig.~\ref{fig:klchapkol}d.


The qualitative features of the time-dependence of $\mathcal{D}^{\rm M}_{q_0}(t)$
are similar to those observed in Fig.~\ref{fig:klchapkol}a -- memory
builds up in a finite interval and smoothly returns back to zero from
the attained maximum. The intuition behind this result is that it
takes a finite time to allow for distinct evolutions of hidden
degrees of freedom that introduce memory in the dynamics of the
reaction coordinate $q_t$. At long times memory is
progressively lost as a result of the gradual relaxation of the hidden
degrees of freedom to their respective equilibrium that in turn
renders the dynamics of the reaction coordinate effectively memory-less
and correspondingly $\mathcal{D}^{\rm M}_{q_0}(t)$ vanishes.

\emph{Single-molecule experiments on a DNA hairpin.---}
As a second example we consider a time-series 
of the end-to-end distance of a single-strand DNA hairpin measured in an optical
tweezers experiment performed by the Woodside group
\cite{neupane_transition-path_2015}. The  data-set contains 11 million
measurements of the extension of the DNA hairpin 30R50T4 held in a
pair of optical traps with
stiffness $0.63$~pN/nm and $1.1$~pN/nm, respectively, sampled with a
$2.5 \mu$s temporal resolution. It has been shown that this time-series is
non-Markovian \cite{pyo_memory_2019}. 
The length of the
time-series 
is
much larger that the relaxation time (see Fig.~\ref{fig:autocorrelation}b) and therefore
we slice it into several pieces that are statistically
independent. More precisely, we use
the time-scale $t_{\rm cut}$ where the
autocorrelation function of the extension, $C(t)$, falls to
$\simeq$0.05. This ensures $t_{\rm cut}\gg t_{\rm rel}$ and yields an ensemble of 50 statistically independent
trajectories. 

We determine the equilibrium probability density $p_{\rm eq}(q)$ (see
inset of Fig.~\ref{fig:autocorrelation}b) and
two-point joint probability density $p(q,t,q_0,0)=p(q,t_0+t,q_0,t_0)$ by performing a
standard histogram analysis with a bin-size of $l_{\rm bin}=$0.35 nm, such that $q$
refers to a bin of width $l_{\rm bin}$ centered at $q$. The Greens
function is thereupon obtained by the law of conditional probability,
$G(q,t|q_0)=p(q,t,q_0,0)/p_{\rm eq}(q)$ while $C(t)$ 
in Eq.~(\ref{autocorr})  is
determined directly from the respective second lines of Eq.~(\ref{exp}).

The Chapman-Kolmogorov construct is determined from $G(q,t|q_0)$ by
direct integration of Eq.~(\ref{CKol}) and is used to determine
$\mathcal{D}^{\rm CK}_{\tau,q_0}(t)$, while the corresponding
fictitious Markovian process evolves as Markovian diffusion in a free energy landscape
$w(q)$ with a constant diffusion coefficient
$D$ \textcolor{black}{that we determine according to standard methods as detailed in the \cite{Note1}.
According to the results 
to a good approximation $D$ is independent of $q$.
The analysis yields $D=447\pm 9$ $\text{nm}^2/\text{ms}$ that we use}
to generate the Markovian
time-series $q_t^{\rm M}$ by integrating the It\^o Langevin equation
(\ref{Langevin}) using the Euler-Mayurama scheme \textcolor{black}{(for details see \cite{Note1})}, and determine $\mathcal{D}^{\rm M}_t(q_0)$ in
Eq.~(\ref{klddiv}) and $C_M(t)$ in Eq.~(\ref{autocorr}),
respectively. 

In contrast to the Rouse-polymer the DNA hairpin exists in two
characteristic conformational states -- folded and unfolded. As a
result, the equilibrium probability density function $p_{\rm eq}(q)$
is bimodal and the dynamics of $q_t$ displays signatures of
metastability \cite{neupane_transition-path_2015}. However, since the
two peaks corresponding to the two sub-populations are not separated
(see inset of Fig.\ref{fig:autocorrelation}b) the potential of mean
force $w(q)$ is expected to underestimate the free energy barrier and
therefore the Markovian evolution is likely to overestimate the relaxation rate. In
complete agreement  Fig.\ref{fig:autocorrelation}b displays an
overestimation of the rate of decay of autocorrelations in the
Markovian approximation by two orders of magnitude in time. Moreover, a long-lived plateau is observed in
the true $C(t)$ spanning more than an order of magnitude in time. 

In order to assess whether the mismatch between true and Markovian
time evolution is predominantly due to an underestimation of the free
energy barrier between folded and unfolded states of the hairpin we inspect the Kullback-Leibler divergence (\ref{klddiv}) between the true and
``Chapman-Kolmogorov evolution'' 
shown in Fig.~\ref{fig:klchapkol}b. The result clearly shows pronounced
signatures of memory 
extending over more than $\sim$10 ms. Note that the ``Chapman-Kolmogorov
evolution'' \textcolor{black}{is exact}
until time
$t=\tau$ whereupon
memory is reset to zero. Therefore a non-zero $\mathcal{D}^{\rm CK}_{\tau,q_0}(t)$ is a clear
signature of memory arising from the dynamical coupling of $q_t$ to
hidden degrees of freedom. Similar to the Rouse-polymer
$\mathcal{D}^{\rm CK}_{\tau,q_0}(t)$ depends on the initial condition
$q_0$. 

A build-up and decay of memory similar to the Rouse-polymer is also
observed in the time evolution of $\mathcal{D}^{\rm M}_{q_0}(t)$, the Kullback-Leibler divergence between
the Green's function of the true evolution and the white-noise
Markovian diffusion in the exact free energy landscape shown in
Fig.~\ref{fig:klchapkol}b. Notably, 
Fig.~\ref{fig:klchapkol}b and Fig.~\ref{fig:klchapkol}e display
essentially the same extent of memory (though the peak is attained
sooner in the white-noise Markovian diffusion), demonstrating that metastability does
not necessarily  destroy nor dominate memory in the evolution of reaction
coordinates. Note that the presence of memory in metastable systems is
not unusual (see e.g. \cite{Dima_2013,Dima_2019} and
\cite{Lapolla_JCP}). In total, the analysis conclusively identifies
extended memory 
in the dynamics of the extension of
the hairpin. 

It is important to note that the extent of memory (of the order of $\sim 10$ms) is
clearly shorter than the relaxation time $t_{\rm rel}$ (compare
Figs.~\ref{fig:autocorrelation}b and \ref{fig:klchapkol}e), and
therefore the decay of memory does not coincide with $t_{\rm rel}$
and the corresponding ``forgetting'' of initial conditions of the
coordinate itself. Instead the memory reflects correlations
between $q_t$ and the \emph{initial conditions} of the hidden degrees of
freedom \cite{lapolla_manifestations_2019}.  The
information encoded in $C(t)$ and $\mathcal{D}^{\rm M,CK}(t)$ is therefore
different --  $\mathcal{D}^{\rm M,CK}(t)$ is a genuine
measure of the extent and duration of memory.

\textcolor{black}{\emph{MD simulation of WW-domain
  of 2F21.---}
  We analyzed 177 atomistic MD
trajectories of the WW-domain of the human
Pin1 Fip (2F21) mutant \cite{jager_structure-function-folding_2006}
provided by the Grubm\"uller group, \emph{each} 1 $\mu$s long sampled every
10 ps. During this time the protein attains a pronounced local equilibrium in the
folded state and does not unfold. The data set was produced in 15 days in ``wall time''. We also
analyzed two longer trajectories, 486 and 651 $\mu$s long sampled
every 200 ps,  from \cite{lindorff-larsen_how_2011}
where the protein reversibly (un)folds several times but
sampling of the unfolded state is limited (see \cite{Note1}).
The fraction of native contacts \cite{best_native_2013} was chosen as
the reaction coordinate (see \cite{Note1} for details). It reflects the displacement of the protein's
structure from the native conformation. In contrast to the previous
examples it is \emph{not} known whether this coordinate displays memory. Technical details
incl. the simulation parameters, estimation of $D(q)$ (with error analysis), and
corresponding results
for the longer trajectories are shown in \cite{Note1}.\\ 
\indent The results are qualitatively similar to the hairpin with one
notable exception --  the diffusion coefficient may not be
considered to 
be constant. The equilibrium density $p_{\rm eq}(q)$ and diffusion
landscape $D(q)$ in the folded state are shown alongside $C(t)$ in
Fig.~\ref{fig:autocorrelation}c. As a first signature of memory the Markovian time-series constructed
according Eq.~\eqref{Langevin} overestimates the relaxation
rate by almost two decades. 
The Kullback-Leibler divergence 
$\mathcal{D}^{\rm CK}_{\tau,q_0}(t)$ in Fig.~\ref{fig:klchapkol}c shows
pronounced memory up to $\sim 10$ ns, extending up to $\sim 100$
ns when considering the longer trajectories that also capture the protein's
dynamics in the unfolded state (see \cite{Note1}). 
Occurring on time-scales $\simeq 20\mu$s
\cite{lindorff-larsen_how_2011}, the (un)folding dynamics is thus memory-less. This example highlights that our method does
\emph{not} distinguish between local and global equilibrium in case of
a time-scale separation, such as the ns time-scale folded-state dynamics and 
$\sim 20\mu$s time-scale (un)folding dynamics.}

\textcolor{black}{
The constructed Markovian time-series shows qualitatively
similar signatures of memory as the hairpin (see
Fig.~\ref{fig:klchapkol}f). The extent of memory displayed by
$\mathcal{D}^{\rm M}_{q_0}(t)$ matches that of $\mathcal{D}^{\rm
  CK}_{\tau,q_0}(t)$ and, 
similar to the Rouse-polymer and hairpin,
depends on the initial condition $q_0$.
One may quite generally relate this dependence
to the dynamics of hidden degrees of freedom with respect to how far $q_0$
is displaced from the free energy minimum.
When $q_0$ is near the free energy minimum the dynamics of hidden degrees of freedom 
has a smaller effect.\\
\indent \emph{Remarks on feasibility.---} The toolbox requires
an ensemble of statistically independent or ergodically long
trajectories. 
Most demanding is the 
Chapman-Kolmogorov analysis that requires sufficient sampling of
the support of the integral in Eq.~\eqref{CKol} at different times
$t,\tau$.
Constructing the Markovian time-series requires 
accurate estimates of $p_{\rm eq}(q)$ and $D(q)$. The
minimal data requirements depend on the system at hand, and may vary
substantially. However, we propose a simple test of the reliability of
the results
--  determining their uncertainty
by a comparison 
with results obtained by
omitting say $\sim$10\%-20\% of data as shown in
Fig.~\ref{fig:klchapkol}e-f. For a reliable quantification of memory
the statistical uncertainty should be substantially smaller than the value of the
Kullback-Leibler divergence,
as in the present case.}

\emph{Conclusion.---} We presented a set of complementary methods to quantify
conclusively the degree and duration of memory in a time series of a reaction coordinate
$q_t$. The proposed toolbox does not assume any particular physical
model. Instead it exploits the Chapman-Kolmogorov equation  
and constructs a fictitious Markovian diffusion process in
the free energy landscape of $q_t$, and compares the artificially
constructed transition probability density with the observed
probability density. The analysis not only determines whether the
dynamics of $q_t$ has memory but also quantifies the magnitude and
duration of memory and thus complements the recently proposed ``test
for Markovianity'' based on transition paths
\cite{berezhkovskii_single-molecule_2018}. 
Whereas in our examples we considered only one-dimensional coordinates, the toolbox generalizes straightforwardly to higher-dimensional reaction
coordinates. The method is general, robust,
and easy to use, \textcolor{black}{and should be used before any attempt to describe a complex system with a low-dimensional Markovian reaction coordinate}. We therefore hope that it will find numerous
applications involving time-series derived from experiments and computer simulations.
 
\paragraph*{Acknowledgments}
We thank Krishna Neupane and Michael T. Woodside for providing
access to their DNA-hairpin data, \textcolor{black}{
  Andreas Volkhardt and Helmut Grubm\"uller for kindly providing
  unpublished MD trajectories, and the D.E. Shaw group for the two
  long MD trajectories published in \cite{lindorff-larsen_how_2011}.} The financial support from the German Research Foundation (DFG) through the Emmy Noether Program GO 2762/1-1 to AG is gratefully acknowledged.

  

\bibliography{FingerPrintsNonMarkovianity.bib}

\begin{thebibliography}{64}%
\makeatletter
\providecommand \@ifxundefined [1]{%
 \@ifx{#1\undefined}
}%
\providecommand \@ifnum [1]{%
 \ifnum #1\expandafter \@firstoftwo
 \else \expandafter \@secondoftwo
 \fi
}%
\providecommand \@ifx [1]{%
 \ifx #1\expandafter \@firstoftwo
 \else \expandafter \@secondoftwo
 \fi
}%
\providecommand \natexlab [1]{#1}%
\providecommand \enquote  [1]{``#1''}%
\providecommand \bibnamefont  [1]{#1}%
\providecommand \bibfnamefont [1]{#1}%
\providecommand \citenamefont [1]{#1}%
\providecommand \href@noop [0]{\@secondoftwo}%
\providecommand \href [0]{\begingroup \@sanitize@url \@href}%
\providecommand \@href[1]{\@@startlink{#1}\@@href}%
\providecommand \@@href[1]{\endgroup#1\@@endlink}%
\providecommand \@sanitize@url [0]{\catcode `\\12\catcode `\$12\catcode
  `\&12\catcode `\#12\catcode `\^12\catcode `\_12\catcode `\%12\relax}%
\providecommand \@@startlink[1]{}%
\providecommand \@@endlink[0]{}%
\providecommand \url  [0]{\begingroup\@sanitize@url \@url }%
\providecommand \@url [1]{\endgroup\@href {#1}{\urlprefix }}%
\providecommand \urlprefix  [0]{URL }%
\providecommand \Eprint [0]{\href }%
\providecommand \doibase [0]{https://doi.org/}%
\providecommand \selectlanguage [0]{\@gobble}%
\providecommand \bibinfo  [0]{\@secondoftwo}%
\providecommand \bibfield  [0]{\@secondoftwo}%
\providecommand \translation [1]{[#1]}%
\providecommand \BibitemOpen [0]{}%
\providecommand \bibitemStop [0]{}%
\providecommand \bibitemNoStop [0]{.\EOS\space}%
\providecommand \EOS [0]{\spacefactor3000\relax}%
\providecommand \BibitemShut  [1]{\csname bibitem#1\endcsname}%
\let\auto@bib@innerbib\@empty
\bibitem [{\citenamefont {Best}\ and\ \citenamefont {Hummer}(2005)}]{Best}%
  \BibitemOpen
  \bibfield  {author} {\bibinfo {author} {\bibfnamefont {R.~B.}\ \bibnamefont
  {Best}}\ and\ \bibinfo {author} {\bibfnamefont {G.}~\bibnamefont {Hummer}},\
  }\bibfield  {title} {\bibinfo {title} {Reaction coordinates and rates from
  transition paths},\ }\href {https://doi.org/10.1073/pnas.0408098102}
  {\bibfield  {journal} {\bibinfo  {journal} {Proc. Natl. Acad. Sci.}\ }\textbf
  {\bibinfo {volume} {102}},\ \bibinfo {pages} {6732–6737} (\bibinfo {year}
  {2005})}\BibitemShut {NoStop}%
\bibitem [{\citenamefont {Portman}\ \emph {et~al.}(2001)\citenamefont
  {Portman}, \citenamefont {Takada},\ and\ \citenamefont {Wolynes}}]{Portman}%
  \BibitemOpen
  \bibfield  {author} {\bibinfo {author} {\bibfnamefont {J.~J.}\ \bibnamefont
  {Portman}}, \bibinfo {author} {\bibfnamefont {S.}~\bibnamefont {Takada}},\
  and\ \bibinfo {author} {\bibfnamefont {P.~G.}\ \bibnamefont {Wolynes}},\
  }\bibfield  {title} {\bibinfo {title} {Microscopic theory of protein folding
  rates. ii. local reaction coordinates and chain dynamics},\ }\href
  {https://doi.org/10.1063/1.1334663} {\bibfield  {journal} {\bibinfo
  {journal} {J. Chem. Phys.}\ }\textbf {\bibinfo {volume} {114}},\ \bibinfo
  {pages} {5082–5096} (\bibinfo {year} {2001})}\BibitemShut {NoStop}%
\bibitem [{\citenamefont {Peters}\ \emph {et~al.}(2013)\citenamefont {Peters},
  \citenamefont {Bolhuis}, \citenamefont {Mullen},\ and\ \citenamefont
  {Shea}}]{Baron}%
  \BibitemOpen
  \bibfield  {author} {\bibinfo {author} {\bibfnamefont {B.}~\bibnamefont
  {Peters}}, \bibinfo {author} {\bibfnamefont {P.~G.}\ \bibnamefont {Bolhuis}},
  \bibinfo {author} {\bibfnamefont {R.~G.}\ \bibnamefont {Mullen}},\ and\
  \bibinfo {author} {\bibfnamefont {J.-E.}\ \bibnamefont {Shea}},\ }\bibfield
  {title} {\bibinfo {title} {Reaction coordinates, one-dimensional smoluchowski
  equations, and a test for dynamical self-consistency},\ }\href
  {https://doi.org/10.1063/1.4775807} {\bibfield  {journal} {\bibinfo
  {journal} {J. Chem. Phys.}\ }\textbf {\bibinfo {volume} {138}},\ \bibinfo
  {pages} {054106} (\bibinfo {year} {2013})}\BibitemShut {NoStop}%
\bibitem [{\citenamefont {Faradjian}\ and\ \citenamefont
  {Elber}(2004)}]{Faradjian}%
  \BibitemOpen
  \bibfield  {author} {\bibinfo {author} {\bibfnamefont {A.~K.}\ \bibnamefont
  {Faradjian}}\ and\ \bibinfo {author} {\bibfnamefont {R.}~\bibnamefont
  {Elber}},\ }\bibfield  {title} {\bibinfo {title} {Computing time scales from
  reaction coordinates by milestoning},\ }\href
  {https://doi.org/10.1063/1.1738640} {\bibfield  {journal} {\bibinfo
  {journal} {J. Chem. Phys.}\ }\textbf {\bibinfo {volume} {120}},\ \bibinfo
  {pages} {10880–10889} (\bibinfo {year} {2004})}\BibitemShut {NoStop}%
\bibitem [{\citenamefont {Berezhkovskii}\ and\ \citenamefont
  {Szabo}(2005)}]{Berezhkovskii_2005}%
  \BibitemOpen
  \bibfield  {author} {\bibinfo {author} {\bibfnamefont {A.}~\bibnamefont
  {Berezhkovskii}}\ and\ \bibinfo {author} {\bibfnamefont {A.}~\bibnamefont
  {Szabo}},\ }\bibfield  {title} {\bibinfo {title} {One-dimensional reaction
  coordinates for diffusive activated rate processes in many dimensions},\
  }\href {https://doi.org/10.1063/1.1818091} {\bibfield  {journal} {\bibinfo
  {journal} {J. Chem. Phys.}\ }\textbf {\bibinfo {volume} {122}},\ \bibinfo
  {pages} {014503} (\bibinfo {year} {2005})}\BibitemShut {NoStop}%
\bibitem [{\citenamefont {Hummer}(2005{\natexlab{a}})}]{Hummer_2005}%
  \BibitemOpen
  \bibfield  {author} {\bibinfo {author} {\bibfnamefont {G.}~\bibnamefont
  {Hummer}},\ }\bibfield  {title} {\bibinfo {title} {Position-dependent
  diffusion coefficients and free energies from bayesian analysis of
  equilibrium and replica molecular dynamics simulations},\ }\href
  {https://doi.org/10.1088/1367-2630/7/1/034} {\bibfield  {journal} {\bibinfo
  {journal} {New J. Phys.}\ }\textbf {\bibinfo {volume} {7}},\ \bibinfo {pages}
  {34–34} (\bibinfo {year} {2005}{\natexlab{a}})}\BibitemShut {NoStop}%
\bibitem [{\citenamefont {Best}\ and\ \citenamefont
  {Hummer}(2009)}]{Best_2009}%
  \BibitemOpen
  \bibfield  {author} {\bibinfo {author} {\bibfnamefont {R.~B.}\ \bibnamefont
  {Best}}\ and\ \bibinfo {author} {\bibfnamefont {G.}~\bibnamefont {Hummer}},\
  }\bibfield  {title} {\bibinfo {title} {Coordinate-dependent diffusion in
  protein folding},\ }\href {https://doi.org/10.1073/pnas.0910390107}
  {\bibfield  {journal} {\bibinfo  {journal} {Proc. Natl. Acad. Sci.}\ }\textbf
  {\bibinfo {volume} {107}},\ \bibinfo {pages} {1088–1093} (\bibinfo {year}
  {2009})}\BibitemShut {NoStop}%
\bibitem [{\citenamefont {Zhang}\ \emph {et~al.}(2016)\citenamefont {Zhang},
  \citenamefont {Hartmann},\ and\ \citenamefont {Schütte}}]{Zhang_2016}%
  \BibitemOpen
  \bibfield  {author} {\bibinfo {author} {\bibfnamefont {W.}~\bibnamefont
  {Zhang}}, \bibinfo {author} {\bibfnamefont {C.}~\bibnamefont {Hartmann}},\
  and\ \bibinfo {author} {\bibfnamefont {C.}~\bibnamefont {Schütte}},\
  }\bibfield  {title} {\bibinfo {title} {Effective dynamics along given
  reaction coordinates, and reaction rate theory},\ }\href
  {https://doi.org/10.1039/c6fd00147e} {\bibfield  {journal} {\bibinfo
  {journal} {Faraday Discussions}\ }\textbf {\bibinfo {volume} {195}},\
  \bibinfo {pages} {365–394} (\bibinfo {year} {2016})}\BibitemShut {NoStop}%
\bibitem [{\citenamefont {Berezhkovskii}\ and\ \citenamefont
  {Makarov}(2017)}]{Dima_2017}%
  \BibitemOpen
  \bibfield  {author} {\bibinfo {author} {\bibfnamefont {A.~M.}\ \bibnamefont
  {Berezhkovskii}}\ and\ \bibinfo {author} {\bibfnamefont {D.~E.}\ \bibnamefont
  {Makarov}},\ }\bibfield  {title} {\bibinfo {title} {Communication:
  Coordinate-dependent diffusivity from single molecule trajectories},\ }\href
  {https://doi.org/10.1063/1.5006456} {\bibfield  {journal} {\bibinfo
  {journal} {J. Chem. Phys.}\ }\textbf {\bibinfo {volume} {147}},\ \bibinfo
  {pages} {201102} (\bibinfo {year} {2017})}\BibitemShut {NoStop}%
\bibitem [{\citenamefont {Berne}\ \emph {et~al.}(1988)\citenamefont {Berne},
  \citenamefont {Borkovec},\ and\ \citenamefont {Straub}}]{Berne_1988}%
  \BibitemOpen
  \bibfield  {author} {\bibinfo {author} {\bibfnamefont {B.~J.}\ \bibnamefont
  {Berne}}, \bibinfo {author} {\bibfnamefont {M.}~\bibnamefont {Borkovec}},\
  and\ \bibinfo {author} {\bibfnamefont {J.~E.}\ \bibnamefont {Straub}},\
  }\bibfield  {title} {\bibinfo {title} {Classical and modern methods in
  reaction rate theory},\ }\href {https://doi.org/10.1021/j100324a007}
  {\bibfield  {journal} {\bibinfo  {journal} {J. Phys. Chem.}\ }\textbf
  {\bibinfo {volume} {92}},\ \bibinfo {pages} {3711–3725} (\bibinfo {year}
  {1988})}\BibitemShut {NoStop}%
\bibitem [{\citenamefont {Hummer}(2005{\natexlab{b}})}]{Hummer}%
  \BibitemOpen
  \bibfield  {author} {\bibinfo {author} {\bibfnamefont {G.}~\bibnamefont
  {Hummer}},\ }\bibfield  {title} {\bibinfo {title} {Position-dependent
  diffusion coefficients and free energies from bayesian analysis of
  equilibrium and replica molecular dynamics simulations},\ }\href
  {https://doi.org/10.1088/1367-2630/7/1/034} {\bibfield  {journal} {\bibinfo
  {journal} {New J. Phys.}\ }\textbf {\bibinfo {volume} {7}},\ \bibinfo {pages}
  {34} (\bibinfo {year} {2005}{\natexlab{b}})}\BibitemShut {NoStop}%
\bibitem [{\citenamefont {Dudko}\ \emph {et~al.}(2008)\citenamefont {Dudko},
  \citenamefont {Hummer},\ and\ \citenamefont {Szabo}}]{Dudko_2008}%
  \BibitemOpen
  \bibfield  {author} {\bibinfo {author} {\bibfnamefont {O.~K.}\ \bibnamefont
  {Dudko}}, \bibinfo {author} {\bibfnamefont {G.}~\bibnamefont {Hummer}},\ and\
  \bibinfo {author} {\bibfnamefont {A.}~\bibnamefont {Szabo}},\ }\bibfield
  {title} {\bibinfo {title} {Theory, analysis, and interpretation of
  single-molecule force spectroscopy experiments},\ }\href
  {https://doi.org/10.1073/pnas.0806085105} {\bibfield  {journal} {\bibinfo
  {journal} {Proc. Natl. Acad. Sci.}\ }\textbf {\bibinfo {volume} {105}},\
  \bibinfo {pages} {15755–15760} (\bibinfo {year} {2008})}\BibitemShut
  {NoStop}%
\bibitem [{\citenamefont {Neupane}\ \emph
  {et~al.}(2016{\natexlab{a}})\citenamefont {Neupane}, \citenamefont {Manuel},\
  and\ \citenamefont {Woodside}}]{neupane_protein_2016}%
  \BibitemOpen
  \bibfield  {author} {\bibinfo {author} {\bibfnamefont {K.}~\bibnamefont
  {Neupane}}, \bibinfo {author} {\bibfnamefont {A.~P.}\ \bibnamefont
  {Manuel}},\ and\ \bibinfo {author} {\bibfnamefont {M.~T.}\ \bibnamefont
  {Woodside}},\ }\bibfield  {title} {\bibinfo {title} {Protein folding
  trajectories can be described quantitatively by one-dimensional diffusion
  over measured energy landscapes},\ }\href {https://doi.org/10.1038/nphys3677}
  {\bibfield  {journal} {\bibinfo  {journal} {Nat. Phys.}\ }\textbf {\bibinfo
  {volume} {12}},\ \bibinfo {pages} {700} (\bibinfo {year}
  {2016}{\natexlab{a}})}\BibitemShut {NoStop}%
\bibitem [{\citenamefont {Neupane}\ \emph
  {et~al.}(2016{\natexlab{b}})\citenamefont {Neupane}, \citenamefont {Foster},
  \citenamefont {Dee}, \citenamefont {Yu}, \citenamefont {Wang},\ and\
  \citenamefont {Woodside}}]{neupane_direct_2016}%
  \BibitemOpen
  \bibfield  {author} {\bibinfo {author} {\bibfnamefont {K.}~\bibnamefont
  {Neupane}}, \bibinfo {author} {\bibfnamefont {D.~A.~N.}\ \bibnamefont
  {Foster}}, \bibinfo {author} {\bibfnamefont {D.~R.}\ \bibnamefont {Dee}},
  \bibinfo {author} {\bibfnamefont {H.}~\bibnamefont {Yu}}, \bibinfo {author}
  {\bibfnamefont {F.}~\bibnamefont {Wang}},\ and\ \bibinfo {author}
  {\bibfnamefont {M.~T.}\ \bibnamefont {Woodside}},\ }\bibfield  {title}
  {\bibinfo {title} {Direct observation of transition paths during the folding
  of proteins and nucleic acids},\ }\href
  {https://doi.org/10.1126/science.aad0637} {\bibfield  {journal} {\bibinfo
  {journal} {Science}\ }\textbf {\bibinfo {volume} {352}},\ \bibinfo {pages}
  {239} (\bibinfo {year} {2016}{\natexlab{b}})}\BibitemShut {NoStop}%
\bibitem [{\citenamefont {Gladrow}\ \emph {et~al.}(2019)\citenamefont
  {Gladrow}, \citenamefont {Ribezzi-Crivellari}, \citenamefont {Ritort},\ and\
  \citenamefont {Keyser}}]{Gladrow}%
  \BibitemOpen
  \bibfield  {author} {\bibinfo {author} {\bibfnamefont {J.}~\bibnamefont
  {Gladrow}}, \bibinfo {author} {\bibfnamefont {M.}~\bibnamefont
  {Ribezzi-Crivellari}}, \bibinfo {author} {\bibfnamefont {F.}~\bibnamefont
  {Ritort}},\ and\ \bibinfo {author} {\bibfnamefont {U.~F.}\ \bibnamefont
  {Keyser}},\ }\bibfield  {title} {\bibinfo {title} {Experimental evidence of
  symmetry breaking of transition-path times},\ }\bibfield  {journal} {\bibinfo
   {journal} {Nat. Commun.}\ }\textbf {\bibinfo {volume} {10}},\ \href
  {https://doi.org/10.1038/s41467-018-07873-9} {10.1038/s41467-018-07873-9}
  (\bibinfo {year} {2019})\BibitemShut {NoStop}%
\bibitem [{\citenamefont {Thorneywork}\ \emph {et~al.}(2020)\citenamefont
  {Thorneywork}, \citenamefont {Gladrow}, \citenamefont {Qing}, \citenamefont
  {Rico-Pasto}, \citenamefont {Ritort}, \citenamefont {Bayley}, \citenamefont
  {Kolomeisky},\ and\ \citenamefont {Keyser}}]{Thorneywork}%
  \BibitemOpen
  \bibfield  {author} {\bibinfo {author} {\bibfnamefont {A.~L.}\ \bibnamefont
  {Thorneywork}}, \bibinfo {author} {\bibfnamefont {J.}~\bibnamefont
  {Gladrow}}, \bibinfo {author} {\bibfnamefont {Y.}~\bibnamefont {Qing}},
  \bibinfo {author} {\bibfnamefont {M.}~\bibnamefont {Rico-Pasto}}, \bibinfo
  {author} {\bibfnamefont {F.}~\bibnamefont {Ritort}}, \bibinfo {author}
  {\bibfnamefont {H.}~\bibnamefont {Bayley}}, \bibinfo {author} {\bibfnamefont
  {A.~B.}\ \bibnamefont {Kolomeisky}},\ and\ \bibinfo {author} {\bibfnamefont
  {U.~F.}\ \bibnamefont {Keyser}},\ }\bibfield  {title} {\bibinfo {title}
  {Direct detection of molecular intermediates from first-passage times},\
  }\bibfield  {journal} {\bibinfo  {journal} {Sci. Adv.}\ }\textbf {\bibinfo
  {volume} {6}},\ \href {https://doi.org/10.1126/sciadv.aaz4642}
  {10.1126/sciadv.aaz4642} (\bibinfo {year} {2020})\BibitemShut {NoStop}%
\bibitem [{\citenamefont {Lapolla}\ and\ \citenamefont
  {Godec}(2019)}]{lapolla_manifestations_2019}%
  \BibitemOpen
  \bibfield  {author} {\bibinfo {author} {\bibfnamefont {A.}~\bibnamefont
  {Lapolla}}\ and\ \bibinfo {author} {\bibfnamefont {A.}~\bibnamefont
  {Godec}},\ }\bibfield  {title} {\bibinfo {title} {Manifestations of
  {Projection}-{Induced} {Memory}: {General} {Theory} and the {Tilted} {Single}
  {File}},\ }\bibfield  {journal} {\bibinfo  {journal} {Front. Phys.}\ }\textbf
  {\bibinfo {volume} {7}},\ \href {https://doi.org/10.3389/fphy.2019.00182}
  {10.3389/fphy.2019.00182} (\bibinfo {year} {2019})\BibitemShut {NoStop}%
\bibitem [{\citenamefont {Hartich}\ and\ \citenamefont
  {Godec}(2020)}]{Hartich_2020}%
  \BibitemOpen
  \bibfield  {author} {\bibinfo {author} {\bibfnamefont {D.}~\bibnamefont
  {Hartich}}\ and\ \bibinfo {author} {\bibfnamefont {A.}~\bibnamefont
  {Godec}},\ }\bibfield  {title} {\bibinfo {title} {Emergent memory and kinetic
  hysteresis in strongly driven networks},\ }\href@noop {} {\bibfield
  {journal} {\bibinfo  {journal} {arXiv:2011.04628}\ } (\bibinfo {year}
  {2020})},\ \Eprint {https://arxiv.org/abs/2011.04628} {arXiv:2011.04628
  [cond-mat.stat-mech]} \BibitemShut {NoStop}%
\bibitem [{\citenamefont {van Kampen}(1998)}]{van_kampen_remarks_1998}%
  \BibitemOpen
  \bibfield  {author} {\bibinfo {author} {\bibfnamefont {N.}~\bibnamefont {van
  Kampen}},\ }\bibfield  {title} {\bibinfo {title} {Remarks on {Non}-{Markov}
  {Processes}},\ }\bibfield  {journal} {\bibinfo  {journal} {Brazilian J.
  Phys.}\ }\textbf {\bibinfo {volume} {28}},\ \href
  {https://doi.org/10.1590/S0103-97331998000200003}
  {10.1590/S0103-97331998000200003} (\bibinfo {year} {1998})\BibitemShut
  {NoStop}%
\bibitem [{\citenamefont {Plotkin}\ and\ \citenamefont
  {Wolynes}(1998)}]{Wolynes}%
  \BibitemOpen
  \bibfield  {author} {\bibinfo {author} {\bibfnamefont {S.~S.}\ \bibnamefont
  {Plotkin}}\ and\ \bibinfo {author} {\bibfnamefont {P.~G.}\ \bibnamefont
  {Wolynes}},\ }\bibfield  {title} {\bibinfo {title} {Non-markovian
  configurational diffusion and reaction coordinates for protein folding},\
  }\href {https://doi.org/10.1103/PhysRevLett.80.5015} {\bibfield  {journal}
  {\bibinfo  {journal} {Phys. Rev. Lett.}\ }\textbf {\bibinfo {volume} {80}},\
  \bibinfo {pages} {5015} (\bibinfo {year} {1998})}\BibitemShut {NoStop}%
\bibitem [{\citenamefont {Zwanzig}(2010)}]{zwanzig_nonequilibrium_2010}%
  \BibitemOpen
  \bibfield  {author} {\bibinfo {author} {\bibfnamefont {R.}~\bibnamefont
  {Zwanzig}},\ }\href@noop {} {\emph {\bibinfo {title} {Nonequilibrium
  statistical mechanics}}}\ (\bibinfo  {publisher} {Oxford Univ. Press},\
  \bibinfo {year} {2010})\BibitemShut {NoStop}%
\bibitem [{\citenamefont {Makarov}(2013)}]{Dima_2013}%
  \BibitemOpen
  \bibfield  {author} {\bibinfo {author} {\bibfnamefont {D.~E.}\ \bibnamefont
  {Makarov}},\ }\bibfield  {title} {\bibinfo {title} {Interplay of non-markov
  and internal friction effects in the barrier crossing kinetics of
  biopolymers: Insights from an analytically solvable model},\ }\href
  {https://doi.org/10.1063/1.4773283} {\bibfield  {journal} {\bibinfo
  {journal} {J. Chem. Phys.}\ }\textbf {\bibinfo {volume} {138}},\ \bibinfo
  {pages} {014102} (\bibinfo {year} {2013})}\BibitemShut {NoStop}%
\bibitem [{\citenamefont {Ozmaian}\ and\ \citenamefont
  {Makarov}(2019)}]{Dima_2019}%
  \BibitemOpen
  \bibfield  {author} {\bibinfo {author} {\bibfnamefont {M.}~\bibnamefont
  {Ozmaian}}\ and\ \bibinfo {author} {\bibfnamefont {D.~E.}\ \bibnamefont
  {Makarov}},\ }\bibfield  {title} {\bibinfo {title} {Transition path dynamics
  in the binding of intrinsically disordered proteins: A simulation study},\
  }\href {https://doi.org/10.1063/1.5129150} {\bibfield  {journal} {\bibinfo
  {journal} {J. Chem. Phys.}\ }\textbf {\bibinfo {volume} {151}},\ \bibinfo
  {pages} {235101} (\bibinfo {year} {2019})}\BibitemShut {NoStop}%
\bibitem [{\citenamefont {Meyer}\ \emph {et~al.}(2020)\citenamefont {Meyer},
  \citenamefont {Pelagejcev},\ and\ \citenamefont {Schilling}}]{Meyer_2020}%
  \BibitemOpen
  \bibfield  {author} {\bibinfo {author} {\bibfnamefont {H.}~\bibnamefont
  {Meyer}}, \bibinfo {author} {\bibfnamefont {P.}~\bibnamefont {Pelagejcev}},\
  and\ \bibinfo {author} {\bibfnamefont {T.}~\bibnamefont {Schilling}},\
  }\bibfield  {title} {\bibinfo {title} {Non-markovian out-of-equilibrium
  dynamics: A general numerical procedure to construct time-dependent memory
  kernels for coarse-grained observables},\ }\href
  {https://doi.org/10.1209/0295-5075/128/40001} {\bibfield  {journal} {\bibinfo
   {journal} {EPL (Europhys. Lett.)}\ }\textbf {\bibinfo {volume} {128}},\
  \bibinfo {pages} {40001} (\bibinfo {year} {2020})}\BibitemShut {NoStop}%
\bibitem [{\citenamefont {Herrera-Delgado}\ \emph {et~al.}(2020)\citenamefont
  {Herrera-Delgado}, \citenamefont {Briscoe},\ and\ \citenamefont
  {Sollich}}]{Sollich_2020}%
  \BibitemOpen
  \bibfield  {author} {\bibinfo {author} {\bibfnamefont {E.}~\bibnamefont
  {Herrera-Delgado}}, \bibinfo {author} {\bibfnamefont {J.}~\bibnamefont
  {Briscoe}},\ and\ \bibinfo {author} {\bibfnamefont {P.}~\bibnamefont
  {Sollich}},\ }\bibfield  {title} {\bibinfo {title} {Tractable nonlinear
  memory functions as a tool to capture and explain dynamical behaviors},\
  }\href {https://doi.org/10.1103/PhysRevResearch.2.043069} {\bibfield
  {journal} {\bibinfo  {journal} {Phys. Rev. Research}\ }\textbf {\bibinfo
  {volume} {2}},\ \bibinfo {pages} {043069} (\bibinfo {year}
  {2020})}\BibitemShut {NoStop}%
\bibitem [{\citenamefont {M\"uller}\ \emph {et~al.}(2020)\citenamefont
  {M\"uller}, \citenamefont {Basu}, \citenamefont {Sollich},\ and\
  \citenamefont {Kr\"uger}}]{Krueger_2020}%
  \BibitemOpen
  \bibfield  {author} {\bibinfo {author} {\bibfnamefont {F.}~\bibnamefont
  {M\"uller}}, \bibinfo {author} {\bibfnamefont {U.}~\bibnamefont {Basu}},
  \bibinfo {author} {\bibfnamefont {P.}~\bibnamefont {Sollich}},\ and\ \bibinfo
  {author} {\bibfnamefont {M.}~\bibnamefont {Kr\"uger}},\ }\bibfield  {title}
  {\bibinfo {title} {Coarse-grained second-order response theory},\ }\href
  {https://doi.org/10.1103/PhysRevResearch.2.043123} {\bibfield  {journal}
  {\bibinfo  {journal} {Phys. Rev. Research}\ }\textbf {\bibinfo {volume}
  {2}},\ \bibinfo {pages} {043123} (\bibinfo {year} {2020})}\BibitemShut
  {NoStop}%
\bibitem [{\citenamefont {Min}\ \emph {et~al.}(2005)\citenamefont {Min},
  \citenamefont {Luo}, \citenamefont {Cherayil}, \citenamefont {Kou},\ and\
  \citenamefont {Xie}}]{Xie_2005}%
  \BibitemOpen
  \bibfield  {author} {\bibinfo {author} {\bibfnamefont {W.}~\bibnamefont
  {Min}}, \bibinfo {author} {\bibfnamefont {G.}~\bibnamefont {Luo}}, \bibinfo
  {author} {\bibfnamefont {B.~J.}\ \bibnamefont {Cherayil}}, \bibinfo {author}
  {\bibfnamefont {S.~C.}\ \bibnamefont {Kou}},\ and\ \bibinfo {author}
  {\bibfnamefont {X.~S.}\ \bibnamefont {Xie}},\ }\bibfield  {title} {\bibinfo
  {title} {Observation of a power-law memory kernel for fluctuations within a
  single protein molecule},\ }\href
  {https://doi.org/10.1103/PhysRevLett.94.198302} {\bibfield  {journal}
  {\bibinfo  {journal} {Phys. Rev. Lett.}\ }\textbf {\bibinfo {volume} {94}},\
  \bibinfo {pages} {198302} (\bibinfo {year} {2005})}\BibitemShut {NoStop}%
\bibitem [{\citenamefont {Lapolla}\ and\ \citenamefont
  {Godec}(2020{\natexlab{a}})}]{Lapolla_2020}%
  \BibitemOpen
  \bibfield  {author} {\bibinfo {author} {\bibfnamefont {A.}~\bibnamefont
  {Lapolla}}\ and\ \bibinfo {author} {\bibfnamefont {A.}~\bibnamefont
  {Godec}},\ }\bibfield  {title} {\bibinfo {title} {Faster uphill relaxation in
  thermodynamically equidistant temperature quenches},\ }\href
  {https://doi.org/10.1103/PhysRevLett.125.110602} {\bibfield  {journal}
  {\bibinfo  {journal} {Phys. Rev. Lett.}\ }\textbf {\bibinfo {volume} {125}},\
  \bibinfo {pages} {110602} (\bibinfo {year} {2020}{\natexlab{a}})}\BibitemShut
  {NoStop}%
\bibitem [{\citenamefont {Lapolla}\ and\ \citenamefont
  {Godec}(2021)}]{Lapolla_2021}%
  \BibitemOpen
  \bibfield  {author} {\bibinfo {author} {\bibfnamefont {A.}~\bibnamefont
  {Lapolla}}\ and\ \bibinfo {author} {\bibfnamefont {A.}~\bibnamefont
  {Godec}},\ }\bibfield  {title} {\bibinfo {title} {Bethesf: Efficient
  computation of the exact tagged-particle propagator in single-file systems
  via the {B}ethe eigenspectrum},\ }\href
  {https://doi.org/10.1016/j.cpc.2020.107569} {\bibfield  {journal} {\bibinfo
  {journal} {Comput. Phys. Commun.}\ }\textbf {\bibinfo {volume} {258}},\
  \bibinfo {pages} {107569} (\bibinfo {year} {2021})}\BibitemShut {NoStop}%
\bibitem [{\citenamefont {Lapolla}\ and\ \citenamefont
  {Godec}(2020{\natexlab{b}})}]{Lapolla_JCP}%
  \BibitemOpen
  \bibfield  {author} {\bibinfo {author} {\bibfnamefont {A.}~\bibnamefont
  {Lapolla}}\ and\ \bibinfo {author} {\bibfnamefont {A.}~\bibnamefont
  {Godec}},\ }\bibfield  {title} {\bibinfo {title} {Single-file diffusion in a
  bi-stable potential: Signatures of memory in the barrier-crossing of a
  tagged-particle},\ }\href {https://doi.org/10.1063/5.0025785} {\bibfield
  {journal} {\bibinfo  {journal} {J. Chem. Phys.}\ }\textbf {\bibinfo {volume}
  {153}},\ \bibinfo {pages} {194104} (\bibinfo {year}
  {2020}{\natexlab{b}})}\BibitemShut {NoStop}%
\bibitem [{\citenamefont {Lapolla}\ and\ \citenamefont
  {Godec}(2018)}]{Lapolla_2018}%
  \BibitemOpen
  \bibfield  {author} {\bibinfo {author} {\bibfnamefont {A.}~\bibnamefont
  {Lapolla}}\ and\ \bibinfo {author} {\bibfnamefont {A.}~\bibnamefont
  {Godec}},\ }\bibfield  {title} {\bibinfo {title} {Unfolding tagged particle
  histories in single-file diffusion: exact single- and two-tag local times
  beyond large deviation theory},\ }\href
  {https://doi.org/10.1088/1367-2630/aaea1b} {\bibfield  {journal} {\bibinfo
  {journal} {New J. Phys.}\ }\textbf {\bibinfo {volume} {20}},\ \bibinfo
  {pages} {113021} (\bibinfo {year} {2018})}\BibitemShut {NoStop}%
\bibitem [{\citenamefont {Kou}\ and\ \citenamefont {Xie}(2004)}]{Xie_2004}%
  \BibitemOpen
  \bibfield  {author} {\bibinfo {author} {\bibfnamefont {S.~C.}\ \bibnamefont
  {Kou}}\ and\ \bibinfo {author} {\bibfnamefont {X.~S.}\ \bibnamefont {Xie}},\
  }\bibfield  {title} {\bibinfo {title} {Generalized {L}angevin equation with
  fractional gaussian noise: Subdiffusion within a single protein molecule},\
  }\href {https://doi.org/10.1103/PhysRevLett.93.180603} {\bibfield  {journal}
  {\bibinfo  {journal} {Phys. Rev. Lett.}\ }\textbf {\bibinfo {volume} {93}},\
  \bibinfo {pages} {180603} (\bibinfo {year} {2004})}\BibitemShut {NoStop}%
\bibitem [{\citenamefont {Neusius}\ \emph {et~al.}(2008)\citenamefont
  {Neusius}, \citenamefont {Daidone}, \citenamefont {Sokolov},\ and\
  \citenamefont {Smith}}]{JCS_2008}%
  \BibitemOpen
  \bibfield  {author} {\bibinfo {author} {\bibfnamefont {T.}~\bibnamefont
  {Neusius}}, \bibinfo {author} {\bibfnamefont {I.}~\bibnamefont {Daidone}},
  \bibinfo {author} {\bibfnamefont {I.~M.}\ \bibnamefont {Sokolov}},\ and\
  \bibinfo {author} {\bibfnamefont {J.~C.}\ \bibnamefont {Smith}},\ }\bibfield
  {title} {\bibinfo {title} {Subdiffusion in peptides originates from the
  fractal-like structure of configuration space},\ }\href
  {https://doi.org/10.1103/PhysRevLett.100.188103} {\bibfield  {journal}
  {\bibinfo  {journal} {Phys. Rev. Lett.}\ }\textbf {\bibinfo {volume} {100}},\
  \bibinfo {pages} {188103} (\bibinfo {year} {2008})}\BibitemShut {NoStop}%
\bibitem [{\citenamefont {Pressé}\ \emph {et~al.}(2014)\citenamefont
  {Pressé}, \citenamefont {Peterson}, \citenamefont {Lee}, \citenamefont
  {Elms}, \citenamefont {MacCallum}, \citenamefont {Marqusee}, \citenamefont
  {Bustamante},\ and\ \citenamefont {Dill}}]{Dill_2014}%
  \BibitemOpen
  \bibfield  {author} {\bibinfo {author} {\bibfnamefont {S.}~\bibnamefont
  {Pressé}}, \bibinfo {author} {\bibfnamefont {J.}~\bibnamefont {Peterson}},
  \bibinfo {author} {\bibfnamefont {J.}~\bibnamefont {Lee}}, \bibinfo {author}
  {\bibfnamefont {P.}~\bibnamefont {Elms}}, \bibinfo {author} {\bibfnamefont
  {J.~L.}\ \bibnamefont {MacCallum}}, \bibinfo {author} {\bibfnamefont
  {S.}~\bibnamefont {Marqusee}}, \bibinfo {author} {\bibfnamefont
  {C.}~\bibnamefont {Bustamante}},\ and\ \bibinfo {author} {\bibfnamefont
  {K.}~\bibnamefont {Dill}},\ }\bibfield  {title} {\bibinfo {title} {Single
  molecule conformational memory extraction: P5ab rna hairpin},\ }\href
  {https://doi.org/10.1021/jp500611f} {\bibfield  {journal} {\bibinfo
  {journal} {J. Phys. Chem. B}\ }\textbf {\bibinfo {volume} {118}},\ \bibinfo
  {pages} {6597–6603} (\bibinfo {year} {2014})}\BibitemShut {NoStop}%
\bibitem [{\citenamefont {Hu}\ \emph {et~al.}(2015)\citenamefont {Hu},
  \citenamefont {Hong}, \citenamefont {Dean~Smith}, \citenamefont {Neusius},
  \citenamefont {Cheng},\ and\ \citenamefont {Smith}}]{Hu_2015}%
  \BibitemOpen
  \bibfield  {author} {\bibinfo {author} {\bibfnamefont {X.}~\bibnamefont
  {Hu}}, \bibinfo {author} {\bibfnamefont {L.}~\bibnamefont {Hong}}, \bibinfo
  {author} {\bibfnamefont {M.}~\bibnamefont {Dean~Smith}}, \bibinfo {author}
  {\bibfnamefont {T.}~\bibnamefont {Neusius}}, \bibinfo {author} {\bibfnamefont
  {X.}~\bibnamefont {Cheng}},\ and\ \bibinfo {author} {\bibfnamefont
  {J.}~\bibnamefont {Smith}},\ }\bibfield  {title} {\bibinfo {title} {The
  dynamics of single protein molecules is non-equilibrium and self-similar over
  thirteen decades in time},\ }\href {https://doi.org/10.1038/nphys3553}
  {\bibfield  {journal} {\bibinfo  {journal} {Nature Physics}\ }\textbf
  {\bibinfo {volume} {12}},\ \bibinfo {pages} {171–174} (\bibinfo {year}
  {2015})}\BibitemShut {NoStop}%
\bibitem [{\citenamefont {Sangha}\ and\ \citenamefont
  {Keyes}(2009)}]{sangha_proteins_2009}%
  \BibitemOpen
  \bibfield  {author} {\bibinfo {author} {\bibfnamefont {A.~K.}\ \bibnamefont
  {Sangha}}\ and\ \bibinfo {author} {\bibfnamefont {T.}~\bibnamefont {Keyes}},\
  }\bibfield  {title} {\bibinfo {title} {Proteins {Fold} by {Subdiffusion} of
  the {Order} {Parameter}},\ }\href {https://doi.org/10.1021/jp907009r}
  {\bibfield  {journal} {\bibinfo  {journal} {J. Phys. Chem. B}\ }\textbf
  {\bibinfo {volume} {113}},\ \bibinfo {pages} {15886} (\bibinfo {year}
  {2009})}\BibitemShut {NoStop}%
\bibitem [{\citenamefont {Avdoshenko}\ \emph {et~al.}(2017)\citenamefont
  {Avdoshenko}, \citenamefont {Das}, \citenamefont {Satija}, \citenamefont
  {Papoian},\ and\ \citenamefont {Makarov}}]{avdoshenko_theoretical_2017}%
  \BibitemOpen
  \bibfield  {author} {\bibinfo {author} {\bibfnamefont {S.~M.}\ \bibnamefont
  {Avdoshenko}}, \bibinfo {author} {\bibfnamefont {A.}~\bibnamefont {Das}},
  \bibinfo {author} {\bibfnamefont {R.}~\bibnamefont {Satija}}, \bibinfo
  {author} {\bibfnamefont {G.~A.}\ \bibnamefont {Papoian}},\ and\ \bibinfo
  {author} {\bibfnamefont {D.~E.}\ \bibnamefont {Makarov}},\ }\bibfield
  {title} {\bibinfo {title} {Theoretical and computational validation of the
  {Kuhn} barrier friction mechanism in unfolded proteins},\ }\href
  {https://doi.org/10.1038/s41598-017-00287-5} {\bibfield  {journal} {\bibinfo
  {journal} {Sci. Rep.}\ }\textbf {\bibinfo {volume} {7}},\ \bibinfo {pages}
  {269} (\bibinfo {year} {2017})}\BibitemShut {NoStop}%
\bibitem [{\citenamefont {Cote}\ \emph {et~al.}(2012)\citenamefont {Cote},
  \citenamefont {Senet}, \citenamefont {Delarue}, \citenamefont {Maisuradze},\
  and\ \citenamefont {Scheraga}}]{cote_anomalous_2012}%
  \BibitemOpen
  \bibfield  {author} {\bibinfo {author} {\bibfnamefont {Y.}~\bibnamefont
  {Cote}}, \bibinfo {author} {\bibfnamefont {P.}~\bibnamefont {Senet}},
  \bibinfo {author} {\bibfnamefont {P.}~\bibnamefont {Delarue}}, \bibinfo
  {author} {\bibfnamefont {G.~G.}\ \bibnamefont {Maisuradze}},\ and\ \bibinfo
  {author} {\bibfnamefont {H.~A.}\ \bibnamefont {Scheraga}},\ }\bibfield
  {title} {\bibinfo {title} {Anomalous diffusion and dynamical correlation
  between the side chains and the main chain of proteins in their native
  state},\ }\href {https://doi.org/10.1073/pnas.1207083109} {\bibfield
  {journal} {\bibinfo  {journal} {Proc. Natl. Acad. Sci.}\ }\textbf {\bibinfo
  {volume} {109}},\ \bibinfo {pages} {10346} (\bibinfo {year}
  {2012})}\BibitemShut {NoStop}%
\bibitem [{\citenamefont {Grossman-Haham}\ \emph {et~al.}(2018)\citenamefont
  {Grossman-Haham}, \citenamefont {Rosenblum}, \citenamefont {Namani},\ and\
  \citenamefont {Hofmann}}]{grossman-haham_slow_2018}%
  \BibitemOpen
  \bibfield  {author} {\bibinfo {author} {\bibfnamefont {I.}~\bibnamefont
  {Grossman-Haham}}, \bibinfo {author} {\bibfnamefont {G.}~\bibnamefont
  {Rosenblum}}, \bibinfo {author} {\bibfnamefont {T.}~\bibnamefont {Namani}},\
  and\ \bibinfo {author} {\bibfnamefont {H.}~\bibnamefont {Hofmann}},\
  }\bibfield  {title} {\bibinfo {title} {Slow domain reconfiguration causes
  power-law kinetics in a two-state enzyme},\ }\href
  {https://doi.org/10.1073/pnas.1714401115} {\bibfield  {journal} {\bibinfo
  {journal} {Proc. Natl. Acad. Sci.}\ }\textbf {\bibinfo {volume} {115}},\
  \bibinfo {pages} {513} (\bibinfo {year} {2018})}\BibitemShut {NoStop}%
\bibitem [{\citenamefont {Pyo}\ and\ \citenamefont
  {Woodside}(2019)}]{pyo_memory_2019}%
  \BibitemOpen
  \bibfield  {author} {\bibinfo {author} {\bibfnamefont {A.~G.~T.}\
  \bibnamefont {Pyo}}\ and\ \bibinfo {author} {\bibfnamefont {M.~T.}\
  \bibnamefont {Woodside}},\ }\bibfield  {title} {\bibinfo {title} {Memory
  effects in single-molecule force spectroscopy measurements of biomolecular
  folding},\ }\href {https://doi.org/10.1039/C9CP04197D} {\bibfield  {journal}
  {\bibinfo  {journal} {Phys. Chem. Chem. Phys.}\ }\textbf {\bibinfo {volume}
  {21}},\ \bibinfo {pages} {24527} (\bibinfo {year} {2019})}\BibitemShut
  {NoStop}%
\bibitem [{\citenamefont {Lu}(1998)}]{enzyme}%
  \BibitemOpen
  \bibfield  {author} {\bibinfo {author} {\bibfnamefont {H.~P.}\ \bibnamefont
  {Lu}},\ }\bibfield  {title} {\bibinfo {title} {Single-molecule enzymatic
  dynamics},\ }\href {https://doi.org/10.1126/science.282.5395.1877} {\bibfield
   {journal} {\bibinfo  {journal} {Science}\ }\textbf {\bibinfo {volume}
  {282}},\ \bibinfo {pages} {1877–1882} (\bibinfo {year} {1998})}\BibitemShut
  {NoStop}%
\bibitem [{\citenamefont {English}\ \emph {et~al.}(2005)\citenamefont
  {English}, \citenamefont {Min}, \citenamefont {van Oijen}, \citenamefont
  {Lee}, \citenamefont {Luo}, \citenamefont {Sun}, \citenamefont {Cherayil},
  \citenamefont {Kou},\ and\ \citenamefont {Xie}}]{enzyme_2}%
  \BibitemOpen
  \bibfield  {author} {\bibinfo {author} {\bibfnamefont {B.~P.}\ \bibnamefont
  {English}}, \bibinfo {author} {\bibfnamefont {W.}~\bibnamefont {Min}},
  \bibinfo {author} {\bibfnamefont {A.~M.}\ \bibnamefont {van Oijen}}, \bibinfo
  {author} {\bibfnamefont {K.~T.}\ \bibnamefont {Lee}}, \bibinfo {author}
  {\bibfnamefont {G.}~\bibnamefont {Luo}}, \bibinfo {author} {\bibfnamefont
  {H.}~\bibnamefont {Sun}}, \bibinfo {author} {\bibfnamefont {B.~J.}\
  \bibnamefont {Cherayil}}, \bibinfo {author} {\bibfnamefont {S.~C.}\
  \bibnamefont {Kou}},\ and\ \bibinfo {author} {\bibfnamefont {X.~S.}\
  \bibnamefont {Xie}},\ }\bibfield  {title} {\bibinfo {title} {Ever-fluctuating
  single enzyme molecules: Michaelis-menten equation revisited},\ }\href
  {https://doi.org/10.1038/nchembio759} {\bibfield  {journal} {\bibinfo
  {journal} {Nat. Chem. Biol.}\ }\textbf {\bibinfo {volume} {2}},\ \bibinfo
  {pages} {87–94} (\bibinfo {year} {2005})}\BibitemShut {NoStop}%
\bibitem [{\citenamefont {Berezhkovskii}\ and\ \citenamefont
  {Makarov}(2018)}]{berezhkovskii_single-molecule_2018}%
  \BibitemOpen
  \bibfield  {author} {\bibinfo {author} {\bibfnamefont {A.~M.}\ \bibnamefont
  {Berezhkovskii}}\ and\ \bibinfo {author} {\bibfnamefont {D.~E.}\ \bibnamefont
  {Makarov}},\ }\bibfield  {title} {\bibinfo {title} {Single-{Molecule} {Test}
  for {Markovianity} of the {Dynamics} along a {Reaction} {Coordinate}},\
  }\href {https://doi.org/10.1021/acs.jpclett.8b00956} {\bibfield  {journal}
  {\bibinfo  {journal} {J. Phys. Chem. Lett.}\ }\textbf {\bibinfo {volume}
  {9}},\ \bibinfo {pages} {2190} (\bibinfo {year} {2018})}\BibitemShut
  {NoStop}%
\bibitem [{\citenamefont {{S. Kullback}}\ and\ \citenamefont {{R.
  Leibler}}(1951)}]{s._kullback_information_1951}%
  \BibitemOpen
  \bibfield  {author} {\bibinfo {author} {\bibnamefont {{S. Kullback}}}\ and\
  \bibinfo {author} {\bibnamefont {{R. Leibler}}},\ }\bibfield  {title}
  {\bibinfo {title} {On information and sufficiency},\ }\href@noop {}
  {\bibfield  {journal} {\bibinfo  {journal} {Ann. Math. Statist}\ }\textbf
  {\bibinfo {volume} {22}},\ \bibinfo {pages} {79} (\bibinfo {year}
  {1951})}\BibitemShut {NoStop}%
\bibitem [{\citenamefont {{Gardiner,
  C.W.}}(1985)}]{gardiner_c.w._handbook_1985}%
  \BibitemOpen
  \bibfield  {author} {\bibinfo {author} {\bibnamefont {{Gardiner, C.W.}}},\
  }\href@noop {} {\emph {\bibinfo {title} {Handbook of {Stochastic} {Methods}
  for {Physics}, {Chemistry} and {Natural} {Sciences}}}},\ \bibinfo {edition}
  {2nd}\ ed.\ (\bibinfo  {publisher} {Springer-Verlag},\ \bibinfo {year}
  {1985})\BibitemShut {NoStop}%
\bibitem [{\citenamefont {Feller}(1959)}]{feller_non-markovian_1959}%
  \BibitemOpen
  \bibfield  {author} {\bibinfo {author} {\bibfnamefont {W.}~\bibnamefont
  {Feller}},\ }\bibfield  {title} {\bibinfo {title} {Non-{Markovian}
  {Processes} with the {Semigroup} {Property}},\ }\href
  {https://doi.org/10.1214/aoms/1177706110} {\bibfield  {journal} {\bibinfo
  {journal} {The Annals of Mathematical Statistics}\ }\textbf {\bibinfo
  {volume} {30}},\ \bibinfo {pages} {1252} (\bibinfo {year}
  {1959})}\BibitemShut {NoStop}%
\bibitem [{\citenamefont {Klimontovich}(1990)}]{Kli}%
  \BibitemOpen
  \bibfield  {author} {\bibinfo {author} {\bibfnamefont {Y.}~\bibnamefont
  {Klimontovich}},\ }\bibfield  {title} {\bibinfo {title} {Ito, {S}tratonovich
  and kinetic forms of stochastic equations},\ }\href
  {https://doi.org/https://doi.org/10.1016/0378-4371(90)90142-F} {\bibfield
  {journal} {\bibinfo  {journal} {Physica A}\ }\textbf {\bibinfo {volume}
  {163}},\ \bibinfo {pages} {515} (\bibinfo {year} {1990})}\BibitemShut
  {NoStop}%
\bibitem [{Note1()}]{Note1}%
  \BibitemOpen
  \bibinfo {note} {See Supplemental Material at [...] for a discretization of
  the anti-It\^o Langevin equation \protect \textup {\hbox {\mathsurround \z@
  \protect \normalfont (\ignorespaces \ref {Langevin}\unskip \@@italiccorr )}},
  exact results for the Rouse polymer, details about MD simulations, the
  fraction of native contacts, the estimation of the diffusion landscape
  $D(q)$, and a description of the uncertainty quantification.}\BibitemShut
  {Stop}%
\bibitem [{\citenamefont {Berezhkovskii}\ and\ \citenamefont
  {Szabo}(2011)}]{Szabo}%
  \BibitemOpen
  \bibfield  {author} {\bibinfo {author} {\bibfnamefont {A.}~\bibnamefont
  {Berezhkovskii}}\ and\ \bibinfo {author} {\bibfnamefont {A.}~\bibnamefont
  {Szabo}},\ }\bibfield  {title} {\bibinfo {title} {Time scale separation leads
  to position-dependent diffusion along a slow coordinate},\ }\href
  {https://doi.org/10.1063/1.3626215} {\bibfield  {journal} {\bibinfo
  {journal} {J. Chem. Phys.}\ }\textbf {\bibinfo {volume} {135}},\ \bibinfo
  {pages} {074108} (\bibinfo {year} {2011})}\BibitemShut {NoStop}%
\bibitem [{\citenamefont {Sunagawa}\ and\ \citenamefont
  {Doi}(1975)}]{sunagawa_theory_1975}%
  \BibitemOpen
  \bibfield  {author} {\bibinfo {author} {\bibfnamefont {S.}~\bibnamefont
  {Sunagawa}}\ and\ \bibinfo {author} {\bibfnamefont {M.}~\bibnamefont {Doi}},\
  }\bibfield  {title} {\bibinfo {title} {Theory of {Diffusion}-{Controlled}
  {Intrachain} {Reactions} of {Polymers}},\ }\href
  {https://doi.org/10.1295/polymj.7.604} {\bibfield  {journal} {\bibinfo
  {journal} {Polymer J.}\ }\textbf {\bibinfo {volume} {7}},\ \bibinfo {pages}
  {604} (\bibinfo {year} {1975})}\BibitemShut {NoStop}%
\bibitem [{\citenamefont {Majumdar}(1999)}]{Satya1}%
  \BibitemOpen
  \bibfield  {author} {\bibinfo {author} {\bibfnamefont {S.~N.}\ \bibnamefont
  {Majumdar}},\ }\bibfield  {title} {\bibinfo {title} {Persistence in
  nonequilibrium systems},\ }\href {http://www.jstor.org/stable/24102955}
  {\bibfield  {journal} {\bibinfo  {journal} {Curr. Sci.}\ }\textbf {\bibinfo
  {volume} {77}},\ \bibinfo {pages} {370} (\bibinfo {year} {1999})}\BibitemShut
  {NoStop}%
\bibitem [{\citenamefont {Bray}\ \emph {et~al.}(2013)\citenamefont {Bray},
  \citenamefont {Majumdar},\ and\ \citenamefont {Schehr}}]{Satya2}%
  \BibitemOpen
  \bibfield  {author} {\bibinfo {author} {\bibfnamefont {A.~J.}\ \bibnamefont
  {Bray}}, \bibinfo {author} {\bibfnamefont {S.~N.}\ \bibnamefont {Majumdar}},\
  and\ \bibinfo {author} {\bibfnamefont {G.}~\bibnamefont {Schehr}},\
  }\bibfield  {title} {\bibinfo {title} {Persistence and first-passage
  properties in nonequilibrium systems},\ }\href
  {https://doi.org/10.1080/00018732.2013.803819} {\bibfield  {journal}
  {\bibinfo  {journal} {Advances in Physics}\ }\textbf {\bibinfo {volume}
  {62}},\ \bibinfo {pages} {225} (\bibinfo {year} {2013})},\ \Eprint
  {https://arxiv.org/abs/https://doi.org/10.1080/00018732.2013.803819}
  {https://doi.org/10.1080/00018732.2013.803819} \BibitemShut {NoStop}%
\bibitem [{\citenamefont {Bray}\ \emph {et~al.}(1994)\citenamefont {Bray},
  \citenamefont {Derrida},\ and\ \citenamefont {Godr{\'{e}}che}}]{Bray_1994}%
  \BibitemOpen
  \bibfield  {author} {\bibinfo {author} {\bibfnamefont {A.~J.}\ \bibnamefont
  {Bray}}, \bibinfo {author} {\bibfnamefont {B.}~\bibnamefont {Derrida}},\ and\
  \bibinfo {author} {\bibfnamefont {C.}~\bibnamefont {Godr{\'{e}}che}},\
  }\bibfield  {title} {\bibinfo {title} {Non-trivial algebraic decay in a
  soluble model of coarsening},\ }\href
  {https://doi.org/10.1209/0295-5075/27/3/001} {\bibfield  {journal} {\bibinfo
  {journal} {Europhys. Lett. ({EPL})}\ }\textbf {\bibinfo {volume} {27}},\
  \bibinfo {pages} {175} (\bibinfo {year} {1994})}\BibitemShut {NoStop}%
\bibitem [{\citenamefont {Derrida}\ \emph {et~al.}(1995)\citenamefont
  {Derrida}, \citenamefont {Hakim},\ and\ \citenamefont {Pasquier}}]{Derrida}%
  \BibitemOpen
  \bibfield  {author} {\bibinfo {author} {\bibfnamefont {B.}~\bibnamefont
  {Derrida}}, \bibinfo {author} {\bibfnamefont {V.}~\bibnamefont {Hakim}},\
  and\ \bibinfo {author} {\bibfnamefont {V.}~\bibnamefont {Pasquier}},\
  }\bibfield  {title} {\bibinfo {title} {Exact first-passage exponents of 1d
  domain growth: Relation to a reaction-diffusion model},\ }\href
  {https://doi.org/10.1103/PhysRevLett.75.751} {\bibfield  {journal} {\bibinfo
  {journal} {Phys. Rev. Lett.}\ }\textbf {\bibinfo {volume} {75}},\ \bibinfo
  {pages} {751} (\bibinfo {year} {1995})}\BibitemShut {NoStop}%
\bibitem [{\citenamefont {Majumdar}\ and\ \citenamefont
  {Bray}(1998)}]{Satya_c1}%
  \BibitemOpen
  \bibfield  {author} {\bibinfo {author} {\bibfnamefont {S.~N.}\ \bibnamefont
  {Majumdar}}\ and\ \bibinfo {author} {\bibfnamefont {A.~J.}\ \bibnamefont
  {Bray}},\ }\bibfield  {title} {\bibinfo {title} {Persistence with partial
  survival},\ }\href {https://doi.org/10.1103/PhysRevLett.81.2626} {\bibfield
  {journal} {\bibinfo  {journal} {Phys. Rev. Lett.}\ }\textbf {\bibinfo
  {volume} {81}},\ \bibinfo {pages} {2626} (\bibinfo {year}
  {1998})}\BibitemShut {NoStop}%
\bibitem [{\citenamefont {Majumdar}\ \emph {et~al.}(1996)\citenamefont
  {Majumdar}, \citenamefont {Bray}, \citenamefont {Cornell},\ and\
  \citenamefont {Sire}}]{Satya_critical}%
  \BibitemOpen
  \bibfield  {author} {\bibinfo {author} {\bibfnamefont {S.~N.}\ \bibnamefont
  {Majumdar}}, \bibinfo {author} {\bibfnamefont {A.~J.}\ \bibnamefont {Bray}},
  \bibinfo {author} {\bibfnamefont {S.~J.}\ \bibnamefont {Cornell}},\ and\
  \bibinfo {author} {\bibfnamefont {C.}~\bibnamefont {Sire}},\ }\bibfield
  {title} {\bibinfo {title} {Global persistence exponent for nonequilibrium
  critical dynamics},\ }\href {https://doi.org/10.1103/PhysRevLett.77.3704}
  {\bibfield  {journal} {\bibinfo  {journal} {Phys. Rev. Lett.}\ }\textbf
  {\bibinfo {volume} {77}},\ \bibinfo {pages} {3704} (\bibinfo {year}
  {1996})}\BibitemShut {NoStop}%
\bibitem [{\citenamefont {Rouse}(1953)}]{rouse_theory_1953}%
  \BibitemOpen
  \bibfield  {author} {\bibinfo {author} {\bibfnamefont {P.~E.}\ \bibnamefont
  {Rouse}},\ }\bibfield  {title} {\bibinfo {title} {A {Theory} of the {Linear}
  {Viscoelastic} {Properties} of {Dilute} {Solutions} of {Coiling}
  {Polymers}},\ }\href {https://doi.org/10.1063/1.1699180} {\bibfield
  {journal} {\bibinfo  {journal} {J. Chem. Phys.}\ }\textbf {\bibinfo {volume}
  {21}},\ \bibinfo {pages} {1272} (\bibinfo {year} {1953})}\BibitemShut
  {NoStop}%
\bibitem [{\citenamefont {Ahn}\ \emph {et~al.}(1993)\citenamefont {Ahn},
  \citenamefont {Schrag},\ and\ \citenamefont {Lee}}]{ahn_bead-spring_1993}%
  \BibitemOpen
  \bibfield  {author} {\bibinfo {author} {\bibfnamefont {K.~H.}\ \bibnamefont
  {Ahn}}, \bibinfo {author} {\bibfnamefont {J.~L.}\ \bibnamefont {Schrag}},\
  and\ \bibinfo {author} {\bibfnamefont {S.~J.}\ \bibnamefont {Lee}},\
  }\bibfield  {title} {\bibinfo {title} {Bead-spring chain model for the
  dynamics of dilute polymer solutions},\ }\href
  {https://doi.org/10.1016/0377-0257(93)80038-D} {\bibfield  {journal}
  {\bibinfo  {journal} {J. Non-Newton Fluid Mech.}\ }\textbf {\bibinfo {volume}
  {50}},\ \bibinfo {pages} {349} (\bibinfo {year} {1993})}\BibitemShut
  {NoStop}%
\bibitem [{\citenamefont {Neupane}\ \emph {et~al.}(2015)\citenamefont
  {Neupane}, \citenamefont {Manuel}, \citenamefont {Lambert},\ and\
  \citenamefont {Woodside}}]{neupane_transition-path_2015}%
  \BibitemOpen
  \bibfield  {author} {\bibinfo {author} {\bibfnamefont {K.}~\bibnamefont
  {Neupane}}, \bibinfo {author} {\bibfnamefont {A.~P.}\ \bibnamefont {Manuel}},
  \bibinfo {author} {\bibfnamefont {J.}~\bibnamefont {Lambert}},\ and\ \bibinfo
  {author} {\bibfnamefont {M.~T.}\ \bibnamefont {Woodside}},\ }\bibfield
  {title} {\bibinfo {title} {Transition-{Path} {Probability} as a {Test} of
  {Reaction}-{Coordinate} {Quality} {Reveals} {DNA} {Hairpin} {Folding} {Is} a
  {One}-{Dimensional} {Diffusive} {Process}},\ }\href
  {https://doi.org/10.1021/acs.jpclett.5b00176} {\bibfield  {journal} {\bibinfo
   {journal} {J. Phys. Chem. Lett.}\ }\textbf {\bibinfo {volume} {6}},\
  \bibinfo {pages} {1005} (\bibinfo {year} {2015})}\BibitemShut {NoStop}%
\bibitem [{\citenamefont {Jager}\ \emph {et~al.}(2006)\citenamefont {Jager},
  \citenamefont {Zhang}, \citenamefont {Bieschke}, \citenamefont {Nguyen},
  \citenamefont {Dendle}, \citenamefont {Bowman}, \citenamefont {Noel},
  \citenamefont {Gruebele},\ and\ \citenamefont
  {Kelly}}]{jager_structure-function-folding_2006}%
  \BibitemOpen
  \bibfield  {author} {\bibinfo {author} {\bibfnamefont {M.}~\bibnamefont
  {Jager}}, \bibinfo {author} {\bibfnamefont {Y.}~\bibnamefont {Zhang}},
  \bibinfo {author} {\bibfnamefont {J.}~\bibnamefont {Bieschke}}, \bibinfo
  {author} {\bibfnamefont {H.}~\bibnamefont {Nguyen}}, \bibinfo {author}
  {\bibfnamefont {M.}~\bibnamefont {Dendle}}, \bibinfo {author} {\bibfnamefont
  {M.~E.}\ \bibnamefont {Bowman}}, \bibinfo {author} {\bibfnamefont {J.~P.}\
  \bibnamefont {Noel}}, \bibinfo {author} {\bibfnamefont {M.}~\bibnamefont
  {Gruebele}},\ and\ \bibinfo {author} {\bibfnamefont {J.~W.}\ \bibnamefont
  {Kelly}},\ }\bibfield  {title} {\bibinfo {title} {Structure-function-folding
  relationship in a {WW} domain},\ }\href
  {https://doi.org/10.1073/pnas.0600511103} {\bibfield  {journal} {\bibinfo
  {journal} {Proc. Natl. Acad. Sci. USA}\ }\textbf {\bibinfo {volume} {103}},\
  \bibinfo {pages} {10648} (\bibinfo {year} {2006})}\BibitemShut {NoStop}%
\bibitem [{\citenamefont {Lindorff-Larsen}\ \emph {et~al.}(2011)\citenamefont
  {Lindorff-Larsen}, \citenamefont {Piana}, \citenamefont {Dror},\ and\
  \citenamefont {Shaw}}]{lindorff-larsen_how_2011}%
  \BibitemOpen
  \bibfield  {author} {\bibinfo {author} {\bibfnamefont {K.}~\bibnamefont
  {Lindorff-Larsen}}, \bibinfo {author} {\bibfnamefont {S.}~\bibnamefont
  {Piana}}, \bibinfo {author} {\bibfnamefont {R.~O.}\ \bibnamefont {Dror}},\
  and\ \bibinfo {author} {\bibfnamefont {D.~E.}\ \bibnamefont {Shaw}},\
  }\bibfield  {title} {\bibinfo {title} {How {Fast}-{Folding} {Proteins}
  {Fold}},\ }\href {https://doi.org/10.1126/science.1208351} {\bibfield
  {journal} {\bibinfo  {journal} {Science}\ }\textbf {\bibinfo {volume}
  {334}},\ \bibinfo {pages} {517} (\bibinfo {year} {2011})}\BibitemShut
  {NoStop}%
\bibitem [{\citenamefont {Best}\ \emph {et~al.}(2013)\citenamefont {Best},
  \citenamefont {Hummer},\ and\ \citenamefont {Eaton}}]{best_native_2013}%
  \BibitemOpen
  \bibfield  {author} {\bibinfo {author} {\bibfnamefont {R.~B.}\ \bibnamefont
  {Best}}, \bibinfo {author} {\bibfnamefont {G.}~\bibnamefont {Hummer}},\ and\
  \bibinfo {author} {\bibfnamefont {W.~A.}\ \bibnamefont {Eaton}},\ }\bibfield
  {title} {\bibinfo {title} {Native contacts determine protein folding
  mechanisms in atomistic simulations},\ }\href
  {https://doi.org/10.1073/pnas.1311599110} {\bibfield  {journal} {\bibinfo
  {journal} {Proc. Natl. Acad. Sci. USA}\ }\textbf {\bibinfo {volume} {110}},\
  \bibinfo {pages} {17874} (\bibinfo {year} {2013})}\BibitemShut {NoStop}%
\bibitem [{\citenamefont {{Abramowitz, Milton and Stegun, Irene
  A.}}(1964)}]{abramowitz}%
  \BibitemOpen
  \bibfield  {author} {\bibinfo {author} {\bibnamefont {{Abramowitz, Milton and
  Stegun, Irene A.}}},\ }\bibfield  {title} {\bibinfo {title} {Handbook of
  {Mathematical} {Functions} with {Formulas}, {Graphs}, and {Mathematical}
  {Tables}},\ }in\ \href@noop {} {\emph {\bibinfo {booktitle} {Handbook of
  {Mathematical} {Functions} with {Formulas}, {Graphs}, and {Mathematical}
  {Tables}}}}\ (\bibinfo  {publisher} {Dover},\ \bibinfo {address} {New York},\
  \bibinfo {year} {1964})\ \bibinfo {edition} {ninth dover printing, tenth gpo
  printing}\ ed.\BibitemShut {Stop}%
\bibitem [{\citenamefont {Johansson}(2017)}]{johansson_arb_2017}%
  \BibitemOpen
  \bibfield  {author} {\bibinfo {author} {\bibfnamefont {F.}~\bibnamefont
  {Johansson}},\ }\bibfield  {title} {\bibinfo {title} {Arb: {Efficient}
  {Arbitrary}-{Precision} {Midpoint}-{Radius} {Interval} {Arithmetic}},\ }\href
  {https://doi.org/10.1109/TC.2017.2690633} {\bibfield  {journal} {\bibinfo
  {journal} {IEEE Transactions on Computers}\ }\textbf {\bibinfo {volume}
  {66}},\ \bibinfo {pages} {1281} (\bibinfo {year} {2017})}\BibitemShut
  {NoStop}%
\end{thebibliography}%


\clearpage
\newpage
\begin{thebibliography}{2}%
\makeatletter
\providecommand \@ifxundefined [1]{%
 \@ifx{#1\undefined}
}%
\providecommand \@ifnum [1]{%
 \ifnum #1\expandafter \@firstoftwo
 \else \expandafter \@secondoftwo
 \fi
}%
\providecommand \@ifx [1]{%
 \ifx #1\expandafter \@firstoftwo
 \else \expandafter \@secondoftwo
 \fi
}%
\providecommand \natexlab [1]{#1}%
\providecommand \enquote  [1]{``#1''}%
\providecommand \bibnamefont  [1]{#1}%
\providecommand \bibfnamefont [1]{#1}%
\providecommand \citenamefont [1]{#1}%
\providecommand \href@noop [0]{\@secondoftwo}%
\providecommand \href [0]{\begingroup \@sanitize@url \@href}%
\providecommand \@href[1]{\@@startlink{#1}\@@href}%
\providecommand \@@href[1]{\endgroup#1\@@endlink}%
\providecommand \@sanitize@url [0]{\catcode `\\12\catcode `\$12\catcode
  `\&12\catcode `\#12\catcode `\^12\catcode `\_12\catcode `\%12\relax}%
\providecommand \@@startlink[1]{}%
\providecommand \@@endlink[0]{}%
\providecommand \url  [0]{\begingroup\@sanitize@url \@url }%
\providecommand \@url [1]{\endgroup\@href {#1}{\urlprefix }}%
\providecommand \urlprefix  [0]{URL }%
\providecommand \Eprint [0]{\href }%
\providecommand \doibase [0]{https://doi.org/}%
\providecommand \selectlanguage [0]{\@gobble}%
\providecommand \bibinfo  [0]{\@secondoftwo}%
\providecommand \bibfield  [0]{\@secondoftwo}%
\providecommand \translation [1]{[#1]}%
\providecommand \BibitemOpen [0]{}%
\providecommand \bibitemStop [0]{}%
\providecommand \bibitemNoStop [0]{.\EOS\space}%
\providecommand \EOS [0]{\spacefactor3000\relax}%
\providecommand \BibitemShut  [1]{\csname bibitem#1\endcsname}%
\let\auto@bib@innerbib\@empty
\bibitem{kloeden_numerical_1994}
Peter~E. Kloeden, Eckhard Platen, and Henri Schurz.
\newblock {\em Numerical {Solution} of {SDE} {Through} {Computer}
  {Experiments}}.
\newblock (Universitext. Springer Berlin Heidelberg, Berlin, Heidelberg, 1994).

\bibitem [{\citenamefont {Rouse}(1953)}]{Srouse_theory_1953}%
  \BibitemOpen
  \bibfield  {author} {\bibinfo {author} {\bibfnamefont {P.~E.}\ \bibnamefont
  {Rouse}},\ }\bibfield  {title} {\bibinfo {title} {A {Theory} of the {Linear}
  {Viscoelastic} {Properties} of {Dilute} {Solutions} of {Coiling}
  {Polymers}},\ }\href {https://doi.org/10.1063/1.1699180} {\bibfield
  {journal} {\bibinfo  {journal} {J. Chem. Phys.}\ }\textbf {\bibinfo {volume}
  {21}},\ \bibinfo {pages} {1272} (\bibinfo {year} {1953})}\BibitemShut
  {NoStop}%
\bibitem [{\citenamefont {Ahn}\ \emph {et~al.}(1993)\citenamefont {Ahn},
  \citenamefont {Schrag},\ and\ \citenamefont {Lee}}]{Sahn_bead-spring_1993}%
  \BibitemOpen
  \bibfield  {author} {\bibinfo {author} {\bibfnamefont {K.~H.}\ \bibnamefont
  {Ahn}}, \bibinfo {author} {\bibfnamefont {J.~L.}\ \bibnamefont {Schrag}},\
  and\ \bibinfo {author} {\bibfnamefont {S.~J.}\ \bibnamefont {Lee}},\
  }\bibfield  {title} {\bibinfo {title} {Bead-spring chain model for the
  dynamics of dilute polymer solutions},\ }\href
  {https://doi.org/10.1016/0377-0257(93)80038-D} {\bibfield  {journal}
  {\bibinfo  {journal} {J. Non-Newton Fluid Mech.}\ }\textbf {\bibinfo {volume}
  {50}},\ \bibinfo {pages} {349} (\bibinfo {year} {1993})}\BibitemShut
  {NoStop}%
\bibitem [{\citenamefont {Lapolla}\ and\ \citenamefont
  {Godec}(2020{\natexlab{a}})}]{SLapolla_2020}%
  \BibitemOpen
  \bibfield  {author} {\bibinfo {author} {\bibfnamefont {A.}~\bibnamefont
  {Lapolla}}\ and\ \bibinfo {author} {\bibfnamefont {A.}~\bibnamefont
  {Godec}},\ }\bibfield  {title} {\bibinfo {title} {Faster uphill relaxation in
  thermodynamically equidistant temperature quenches},\ }\bibfield  {journal}
  {\bibinfo  {journal} {Phys. Rev. Lett.}\ }\textbf {\bibinfo {volume} {125}},\
  \href {https://doi.org/10.1103/physrevlett.125.110602}
  {10.1103/physrevlett.125.110602} (\bibinfo {year}
  {2020}{\natexlab{a}})\BibitemShut {NoStop}%
\bibitem [{\citenamefont {Sunagawa}\ and\ \citenamefont
  {Doi}(1975)}]{Ssunagawa_theory_1975}%
  \BibitemOpen
  \bibfield  {author} {\bibinfo {author} {\bibfnamefont {S.}~\bibnamefont
  {Sunagawa}}\ and\ \bibinfo {author} {\bibfnamefont {M.}~\bibnamefont {Doi}},\
  }\bibfield  {title} {\bibinfo {title} {Theory of {Diffusion}-{Controlled}
  {Intrachain} {Reactions} of {Polymers}},\ }\href
  {https://doi.org/10.1295/polymj.7.604} {\bibfield  {journal} {\bibinfo
  {journal} {Polymer J.}\ }\textbf {\bibinfo {volume} {7}},\ \bibinfo {pages}
  {604} (\bibinfo {year} {1975})}\BibitemShut {NoStop}%

  
\bibitem [{\citenamefont {{Abramowitz, Milton and Stegun, Irene
  A.}}(1964)}]{abramowitz}%
  \BibitemOpen
  \bibfield  {author} {\bibinfo {author} {\bibnamefont {{Abramowitz, Milton and
  Stegun, Irene A.}}},\ }\bibfield  {title} {\bibinfo {title} {Handbook of
  {Mathematical} {Functions} with {Formulas}, {Graphs}, and {Mathematical}
  {Tables}},\ }in\ \href@noop {} {\emph {\bibinfo {booktitle} {Handbook of
  {Mathematical} {Functions} with {Formulas}, {Graphs}, and {Mathematical}
  {Tables}}}}\ (\bibinfo  {publisher} {Dover},\ \bibinfo {address} {New York},\
  \bibinfo {year} {1964})\ \bibinfo {edition} {ninth dover printing, tenth gpo
  printing}\ ed.\BibitemShut {Stop}%
\bibitem [{\citenamefont {Johansson}(2017)}]{johansson_arb_2017}%
  \BibitemOpen
  \bibfield  {author} {\bibinfo {author} {\bibfnamefont {F.}~\bibnamefont
  {Johansson}},\ }\bibfield  {title} {\bibinfo {title} {Arb: {Efficient}
  {Arbitrary}-{Precision} {Midpoint}-{Radius} {Interval} {Arithmetic}},\ }\href
  {https://doi.org/10.1109/TC.2017.2690633} {\bibfield  {journal} {\bibinfo
  {journal} {IEEE Transactions on Computers}\ }\textbf {\bibinfo {volume}
  {66}},\ \bibinfo {pages} {1281} (\bibinfo {year} {2017})}\BibitemShut
  {NoStop}%
\bibitem [{\citenamefont {Wilemski}\ and\ \citenamefont
  {Fixman}(1974)}]{Swilemski_diffusioncontrolled_1974}%
  \BibitemOpen
  \bibfield  {author} {\bibinfo {author} {\bibfnamefont {G.}~\bibnamefont
  {Wilemski}}\ and\ \bibinfo {author} {\bibfnamefont {M.}~\bibnamefont
  {Fixman}},\ }\bibfield  {title} {\bibinfo {title} {Diffusion‐controlled
  intrachain reactions of polymers. {I} {Theory}},\ }\href
  {https://doi.org/10.1063/1.1681162} {\bibfield  {journal} {\bibinfo
  {journal} {J. Chem. Phys.}\ }\textbf {\bibinfo {volume} {60}},\ \bibinfo
  {pages} {866} (\bibinfo {year} {1974})}\BibitemShut {NoStop}%
\bibitem [{\citenamefont {Lapolla}\ and\ \citenamefont
  {Godec}(2019)}]{Slapolla_manifestations_2019}%
  \BibitemOpen
  \bibfield  {author} {\bibinfo {author} {\bibfnamefont {A.}~\bibnamefont
  {Lapolla}}\ and\ \bibinfo {author} {\bibfnamefont {A.}~\bibnamefont
  {Godec}},\ }\bibfield  {title} {\bibinfo {title} {Manifestations of
  {Projection}-{Induced} {Memory}: {General} {Theory} and the {Tilted} {Single}
  {File}},\ }\bibfield  {journal} {\bibinfo  {journal} {Front. Phys.}\ }\textbf
  {\bibinfo {volume} {7}},\ \href {https://doi.org/10.3389/fphy.2019.00182}
  {10.3389/fphy.2019.00182} (\bibinfo {year} {2019})\BibitemShut
  {NoStop}%
\bibitem{best_coordinate-dependent_2010}
R.~B. Best and G.~Hummer.
\newblock Coordinate-dependent diffusion in protein folding.
\newblock {\em Proc. Natl. Acad. Sci. USA}, \textbf{107}, 1088,
(2010).

\bibitem{jager_structure-function-folding_2006}
M.~Jager, Y.~Zhang, J.~Bieschke, H.~Nguyen, M.~Dendle, M.~E. Bowman, J.~P.
  Noel, M.~Gruebele, and J.~W. Kelly.
\newblock Structure-function-folding relationship in a {WW} domain.
\newblock {\em Proc Natl Acad Sci USA} \textbf{103}, 10648 (2006).

\bibitem{pronk_gromacs_2013}
Sander Pronk, Szilárd Páll, Roland Schulz, Per Larsson, Pär Bjelkmar, Rossen
  Apostolov, Michael~R. Shirts, Jeremy~C. Smith, Peter~M. Kasson, David van~der
  Spoel, Berk Hess, and Erik Lindahl.
\newblock {GROMACS} 4.5: a high-throughput and highly parallel open source
  molecular simulation toolkit.
\newblock {\em Bioinformatics}  \textbf{29}, 845 (2013).

\bibitem{lindorff-larsen_improved_2010}
Kresten Lindorff-Larsen, Stefano Piana, Kim Palmo, Paul Maragakis, John~L.
  Klepeis, Ron~O. Dror, and David~E. Shaw.
\newblock Improved side-chain torsion potentials for the {Amber} {ff99SB}
  protein force field: {Improved} {Protein} {Side}-{Chain} {Potentials}.
\newblock {\em Proteins} \textbf{78}, 1950 (2010).

\bibitem{horn_development_2004}
Hans~W. Horn, William~C. Swope, Jed~W. Pitera, Jeffry~D. Madura, Thomas~J.
  Dick, Greg~L. Hura, and Teresa Head-Gordon.
\newblock Development of an improved four-site water model for biomolecular
  simulations: {TIP4P}-{Ew}.
\newblock {\em J. Chem. Phys.}  \textbf{120}, 9665 (2004).

\bibitem{bussi_canonical_2007}
Giovanni Bussi, Davide Donadio, and Michele Parrinello.
\newblock Canonical sampling through velocity rescaling.
\newblock {\em J. Chem. Phys.}  \textbf{126}, 014101 (2007).

\bibitem{parrinello_polymorphic_1981}
M.~Parrinello and A.~Rahman.
\newblock Polymorphic transitions in single crystals: {A} new molecular
  dynamics method.
\newblock {\em J. Appl. Phys.} \textbf{52}, 7182 (1981).

\bibitem{darden_particle_1993}
Tom Darden, Darrin York, and Lee Pedersen.
\newblock Particle mesh {Ewald}: {An} {Nlog}( {N} ) method for {Ewald} sums in
  large systems.
\newblock {\em J. Chem. Phys.} \textbf{98}, 10089  (1993).

\bibitem{best_native_2013}
R.~B. Best, G.~Hummer, and W.~A. Eaton.
\newblock Native contacts determine protein folding mechanisms in atomistic
  simulations.
\newblock {\em Proc. Natl. Acad. Sci. USA} \textbf{110}, 17874 (2013).

\bibitem{mcgibbon_mdtraj_2015}
Robert~T. McGibbon, Kyle~A. Beauchamp, Matthew~P. Harrigan, Christoph Klein,
  Jason~M. Swails, Carlos~X. Hernández, Christian~R. Schwantes, Lee-Ping Wang,
  Thomas~J. Lane, and Vijay~S. Pande.
\newblock {MDTraj}: {A} {Modern} {Open} {Library} for the {Analysis} of
  {Molecular} {Dynamics} {Trajectories}.
\newblock {\em Biophys. J.} \textbf{109}, 1528 (2015).

\bibitem{best_native_2013}
R.~B. Best, G.~Hummer, and W.~A. Eaton.
\newblock Native contacts determine protein folding mechanisms in atomistic
  simulations.
\newblock {\em Proc. Natl. Acad. Sci. USA} \textbf{110}, 17874 (2013).

\bibitem{piana_computational_2011}
Stefano Piana, Krishnarjun Sarkar, Kresten Lindorff-Larsen, Minghao Guo, Martin
  Gruebele, and David~E. Shaw.
\newblock Computational {Design} and {Experimental} {Testing} of the
  {Fastest}-{Folding} $\beta$-{Sheet} {Protein}.
\newblock {\em J. Mol. Bio.} \textbf{405}, 43 (2011).

\bibitem{lindorff-larsen_how_2011}
K.~Lindorff-Larsen, S.~Piana, R.~O. Dror, and D.~E. Shaw.
\newblock How {Fast}-{Folding} {Proteins} {Fold}.
\newblock {\em Science}  \textbf{334}, 517 (2011).

\end{thebibliography}
\clearpage
\newpage
\onecolumngrid
\renewcommand{\thefigure}{S\arabic{figure}}
\renewcommand{\theequation}{S\arabic{equation}}
\setcounter{equation}{0}
\begin{center}
  \textbf{Supplementary Material for:\\
  A Toolbox for Quantifying Memory in Dynamics
  Along Reaction Coordinates}\\[0.2cm]
Alessio Lapolla and Alja\v{z} Godec\\
\emph{Mathematical bioPhysics Group, Max Planck Institute for Biophysical Chemistry, 37077 Göttingen, Germany}
\end{center}

\begin{center}
\textbf{Abstract}\\[0.3cm]  
\end{center}  
\begin{quotation}
  In this  Supplementary Material (SM)  we present details about the
  numerical integration of the anti-It\^o Langevin equation,
all exact results
  for the Rouse polymer, MD simulation details and the fraction of
  native contacts -- the reaction coordinate for the protein dynamics
  considered in the Letter, as well as details about the estimation of the
  diffusion landscape $D(q)$. In addition, a description of the
  uncertainty quantification for the Kullback-Leibler divergence and supplementary figures are included showing the various
Green's functions for the Rouse polymer and DNA hairpin.
\end{quotation}

\section{Discretized Langevin equation: Anti-It\^{o} Euler-Mayurama scheme}
The numerical integration of the overdamped Langevin equation with
multiplicative noise (i.e. the diffusion coefficient $D(q_t)$ depends
on the position $q_t$) in dimension 1 in the post-point anti-It\^o
interpretation with integration step $\Delta t$, i.e. $q_t\to
q_{t+\Delta t}$, is carried out as \cite{kloeden_numerical_1994}
\begin{subequations}
  \begin{equation}
    \tilde{q}_t\equiv q_t+\frac{D(q_t)}{k_{\rm B}T}f(q_t)\Delta t+\sqrt{2D(q_t) \Delta t}\,\eta
  \end{equation}
  \begin{equation}
    q_{q+\Delta t}=\tilde{q}_t+\left(\sqrt{2D(\tilde{q}_t) \Delta t}-\sqrt{2D(q_t) \Delta t}\right)\,\eta,
  \end{equation}
\end{subequations}
where we assumed the validity of the fluctuation-dissipation theorem,
i.e.\ $\mu=D/k_{\rm B}T$ is the mobility,  
$f(q_t)$ is the force, and $\eta$ is a random number drawn from a
Gaussian distribution with zero mean and unit variance. Note that only a
single random number $\eta$ is required for each iteration. When the
noise is additive (i.e. $D(q)\to D$ is a constant) the previous scheme simplifies to the classic Euler-Mayurama scheme
\begin{equation}
  q_{t+\Delta t}=q_t+\frac{D}{k_BT}f(q_t)\Delta t+\sqrt{2D\Delta t}\,\eta.
\end{equation}
The above equations are used to integrate the Langevin equation (5) in
the main text in the case of the DNA hairpin and protein.

\section{Analytical results for the Rouse polymer}
The probability
density function for the positions of all beads $\{\mathbf{r}_i\}$ is
well-known \cite{Srouse_theory_1953, Sahn_bead-spring_1993} and allows
us to determine exactly the probability density of the end-to-end
distance.

Introducing $\nu_k\equiv k\pi/2(N+1)$, $\alpha_k=4\sin^2(\nu_k)$ as well as
$Q_{ik}\equiv\sqrt{2/(N+1)}\cos(\nu_k[2i-1])$ and
$\eta_t\equiv\sum_{k=1}^N(Q_{1k}-Q_{N+1\,k})^2\mathrm{e}^{-\alpha_kt}/2\alpha_k$
the equilibrium probability density of $q$ is
given by $p_{\rm eq}(q)=q^2\mathrm{e}^{-q^2/4\eta_0}/2\sqrt{\pi}\eta_0^{3/2}$
for $q\in[0,\infty)$ with the mean extension $\langle
  d\rangle=4\sqrt{\eta_0/\pi}$ and mean square extension $\langle
  d^2\rangle=6\eta_0$. The probability density function of
  $q$ reads exactly (for a derivation see Ref.~\cite{SLapolla_2020})
 \begin{equation} 
G_R(q,t|q_0)=\frac{q q_0\mathrm{e}^{-\eta_0(q^2+q_0^2)/2(\eta_0^2-\eta_t^2)}}{p_{\rm eq}(q)2\pi
  \eta_t\sqrt{\eta_0^2-\eta_t^2}}\sinh\left(\frac{\eta_tq q_0}{2(\eta_0^2-\eta_t^2)}\right).
   \label{exact_Rouse}
 \end{equation}
The exact autocorrelation function is in turn
obtained in the form
\begin{equation}
C(t)=\frac{6\sqrt{\eta_0^2-\eta_t^2}}{(3\pi-8)\eta_0}+
\frac{4(\eta_0^2+\eta_t^2)\arctan\left(\eta_t/\sqrt{\eta_0^2-\eta_t^2}\right)}{(3\pi-8)\eta_0\eta_t}.
\label{ac_Rouse}  
\end{equation} 
The Fokker-Planck equation in the Markovian approximation
to the
evolution
of $q$ for the Rouse
polymer 
can be solved 
in the form of a spectral expansion
\cite{Ssunagawa_theory_1975} and reads
$G^{\rm M}_R(q,t|q_0)=\sum_{k=0}^\infty\psi_k^R(q)\psi_k^L(q_0)\mathrm{e}^{-kt/\eta_0}$ where 
\begin{equation}
\psi_k^L(x)\equiv\sqrt{\frac{k!\sqrt{\pi}}{2\Gamma(3/2+k)}}L_k^{1/2}\left(\frac{x^2}{4\eta_0}\right)
\end{equation}  
where $\Gamma(x)$ denotes the Gamma-function and $L_k^{1/2}(x)$ the generalized Laguerre polynomial of
degree $k$ with parameter $1/2$ (see \cite{abramowitz}) that we compute
using the Arb-library \cite{johansson_arb_2017} and $\psi_k^R(x)=p_{\rm
  eq}(x)\psi_k^L(x)$. Here from it is straightforward to obtain the
autocorrelation function in the Markovian approximation that reads
\begin{equation}
C_M(t)=\frac{\sqrt{\pi}}{3-8/\pi}\sum_{k=1}^{\infty}\mathrm{e}^{-kt/\eta_0}[k!\Gamma(3/2-k)^2\Gamma(3/2+k)]^{-1}.
\label{ac_Rouse_M} 
\end{equation}
The integral defined in Eq.~(4) in the
main text can be solved analytically via a straightforward but tedious
calculation using Eq.~\eqref{exact_Rouse}. The result of the integral
reads exactly 
\begin{eqnarray}
    G^{CK}_{t_1}(q,t|q_0)&=&\frac{\eta_0^{7/2} q}{q_0\eta_{t-\tau}\eta_\tau\sqrt{\pi\Xi_{\tau,t-\tau}}}
    \sinh\left(\frac{q q_0 \eta_0 \eta_\tau \eta_{t-\tau}}{2\Xi_{\tau,t-\tau}}\right)\nonumber\times\\
    &&\exp\left[-\frac{q^2\eta_0}{2\Omega^-_{t-\tau}}\left(1-\frac{\eta_{t-\tau}^2\Omega^{-}_\tau}{2\Xi_{\tau,t-\tau}}\right)
        -\frac{q_0^2}{4\Omega^-_\tau}\left(\frac{\eta_0^2\Omega_\tau^--\eta_{t-\tau}^2\Omega^+_\tau}{\Omega^-_{t-\tau}}-\frac{\eta_0\eta_\tau^2\Omega^{-}_{t-\tau}}{\Xi_{\tau,t-\tau}}\right)\right]
    \label{CKG}
\end{eqnarray}
having defined
\begin{equation}
    \Omega_t^{\pm}=\eta_0^2\pm\eta_t^2,\quad
    \Xi_{\tau,t-\tau}=4\eta_0^4-\Omega_\tau^{+}\Omega_{t-\tau}^{+}.
\end{equation}
Notably, the structure of Eq.~\eqref{CKG} is identical to the
structure of the plain Green's function (Eq.~\eqref{exact_Rouse}) but
here the temporal dependence is obviously different.

Note that in when the observation time is much larger than the
relaxation time of the observable $t_\mathrm{rel}$, we find for
$t-\tau>t_\mathrm{rel}$ that $G_\tau^{\mathrm{CK}}(q,t|q_0)\simeq
p_\mathrm{eq}(q)\int dq' G(q',t|q_0)=p_\mathrm{eq}(q)$. Therefore,
since $\lim_{t\to\infty}G(q,t|q')=p_\mathrm{eq}(q)$, the definition of
$G_\tau^{\mathrm{CK}}(q,t|q_0)$ (Eq.~(4) in the main text) by
construction ensures
$\lim_{t\to\infty}\mathcal{D}^\mathrm{CK}_{\tau,q_0}(t)=0$.

\section{Green's functions}
\begin{figure}
    \centering
    \includegraphics[width=0.7\textwidth]{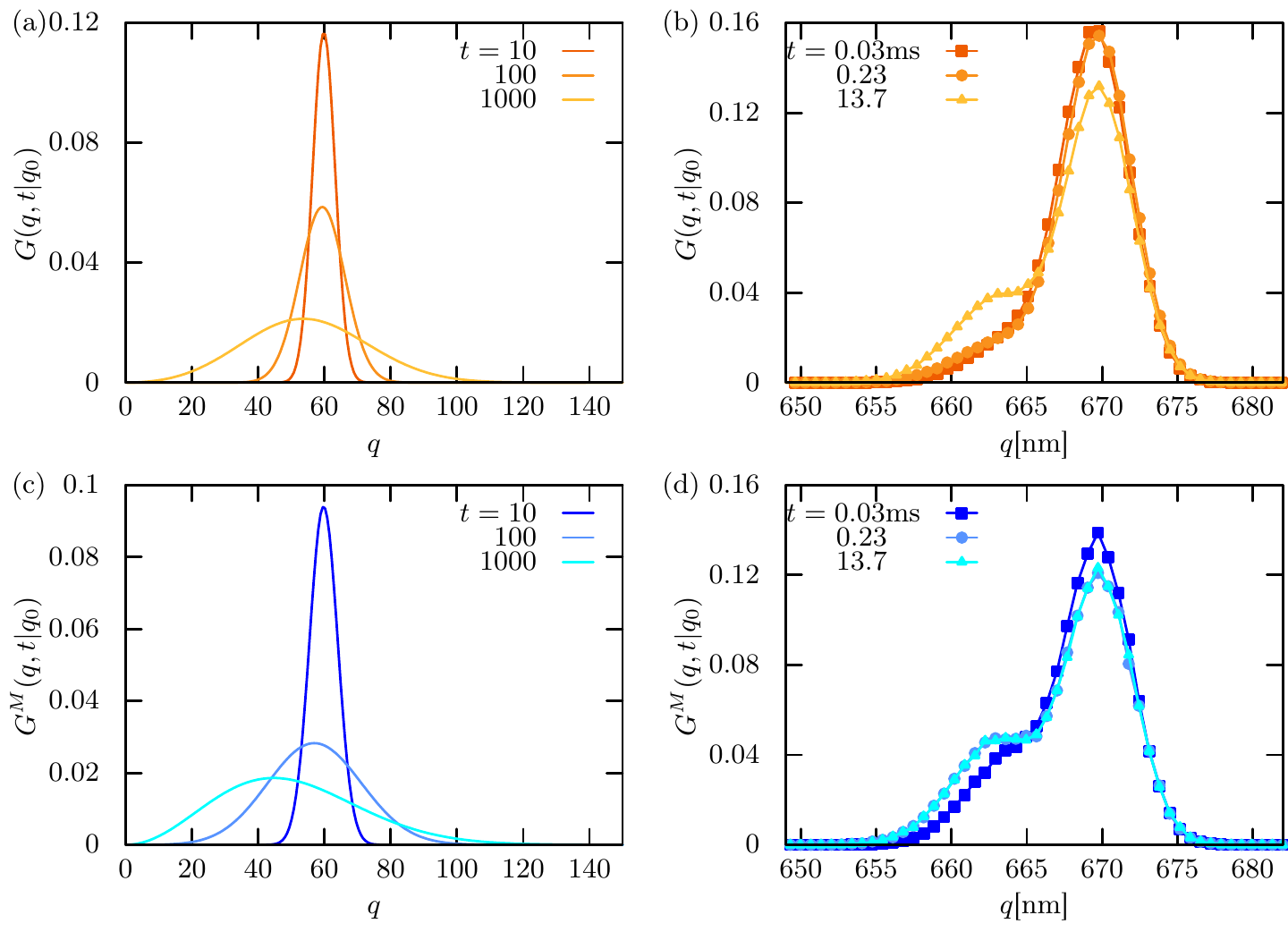}
    \caption{Green's function at different times for both considered
      systems. a) and b) depict the true Green's function for the
      end-to-end distance of the Rouse chain and of the DNA hairpin
      respectively. Panels c) and d) show the Green's function of
      their respective fictitious Markovian processes at the same
      times. The initial conditions are $q_0=60$ for the Rouse chain and
      $q_0=671$ nm for the hairpin.}
    \label{fig:greens}
\end{figure}

In
Fig.~\ref{fig:greens} we explicitly show the Green's function that is required for the computation of the Kullback-Liebler divergence. 

\section{Details of the projection affect the relaxation time and extent of memory}
 In the main text we consider Rouse polymer chain composed of $1000$
 beads and we focus on the autocorrelation function of its end-to-end
 distance as the reaction coordinate $q_t$. We find that the
 fictitious Markovian reference process describing Brownian diffusion
 in the free energy landscape overestimates the relaxation rate; a
 similar observation is also made in the case of the experimental
 hairpin data.
 However this difference in the rate of relaxation is non-unique and
 in fact depends on the  observable, i.e. on details of the projection. 
 \begin{figure}[ht!!]
    \centering
    \includegraphics[width=0.4\textwidth]{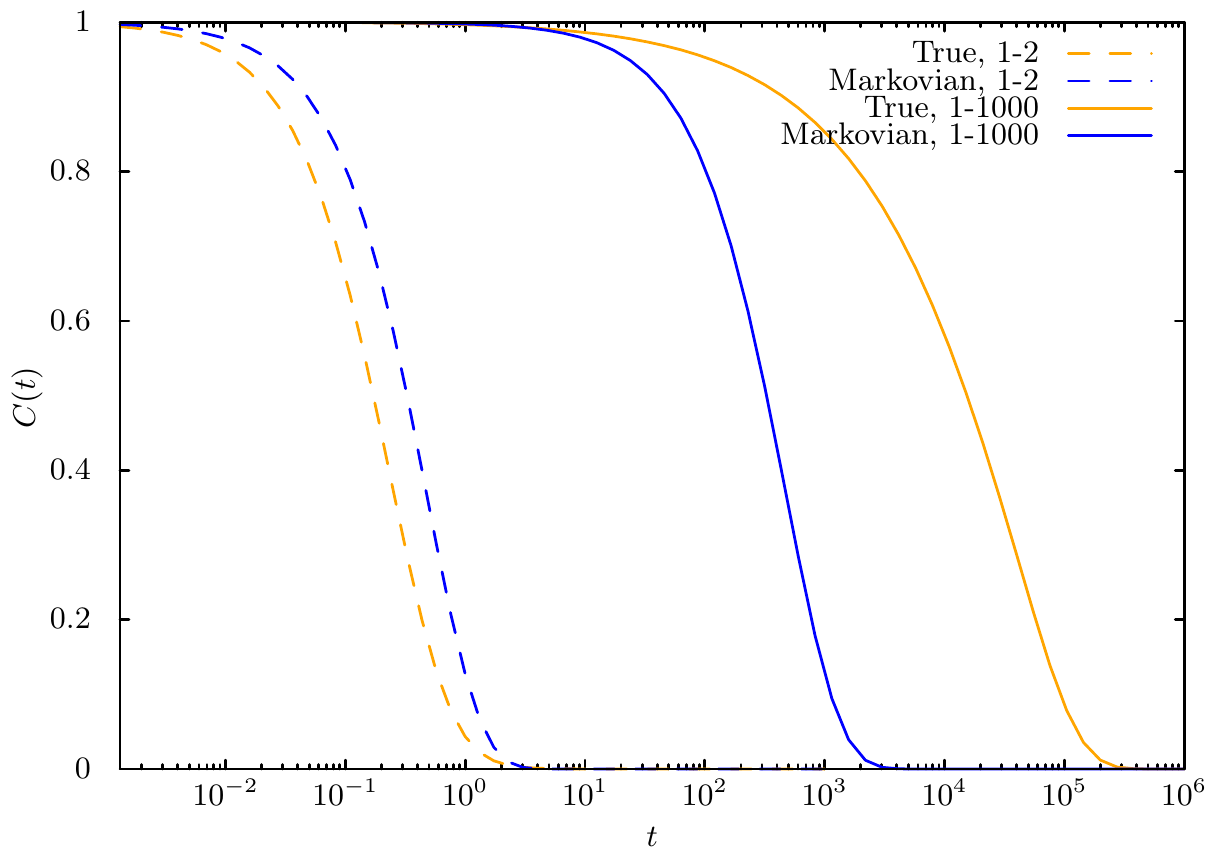}
    \caption{The autocorrelation function of the distance between the
      first and the second bead (dashed lines) and first and last bead (full
      lines) of a Rouse chain composed by $1000$ according to the true
      (orange) and fictitious Markovian evolution (blue). Note the the
      free energy landscape for both orange-blue pairs is by
      construction identical. The
      continuous lines are those shown in the main text.}
    \label{fig:auto12}
 \end{figure}
 
For example we demonstrate in Fig.~\ref{fig:auto12} the opposite trend
that arises when we observe the autocorrelation function of the
distance between the first and the second bead of the same Rouse Chain
(see dashed lines).

In addition, is worth to note that if the Green's
function $\mathcal{G}$ describing the full many-dimensional system is
diagonalizable  (like in the Rouse chain
case~\cite{Swilemski_diffusioncontrolled_1974} or any Markovian
dynamics obeying detailed balance), it can be written as 
\begin{equation}
 \mathcal{G}(\mathbf{x},t|\mathbf{x}_0)=\sum_{k} \psi_k^R(\mathbf{x}) \psi_k^L(\mathbf{x}_0) \mathrm{e}^{-\lambda_k t},
\end{equation}
where $\psi_k^R$ and $\psi_k^L$ are respectively the right and left
eigenfunctions of the underlying Fokker-Planck-Smoluchowski operator,
while $\lambda_k$ denotes the eigenvalues. Then the Green's function
of the projected observable -- the reaction coordinate $q=\Gamma(\mathbf{x})$ -- can be written in full generality~\cite{Slapolla_manifestations_2019} as
\begin{equation}
 G(q,t|q_0)=\sum_{k} V^{R}_k(q;\Gamma)V^{L}_k(q_0;\Gamma)\mathrm{e}^{-\lambda_k t},
\end{equation}
where the elements $V^{R}_k$ and $V^{L}_k$ depend both on $\psi_k^R$
and $\psi_k^L$, and on the projection $\Gamma(\mathbf{x})$. In turn the autocorrelation function can be easily computed as:
\begin{align}
 &C(t)=\sum_{k} (\int q V^{R}_k(q;\Gamma)dq)(\int q_0 V^{L}_k(q_0;\Gamma)dq_0)\mathrm{e}^{-\lambda_k t}\equiv \sum_ka^{R,\Gamma}_k b^{L,\Gamma}_k \mathrm{e}^{-\lambda_k t},
\end{align}
and one can show that for systems obeying detailed balance $a^{R,\Gamma}_k
b^{L,\Gamma}_k \geq 0$ \cite{Slapolla_manifestations_2019}.
The analysis shows that the projection only affects the weights
whereas the exponentiated eigenvalues (and thus time-scales) are those
of the full system's dynamics.

Nevertheless, the autocorrelation function of different observables of
the same system may decay on widely disparate time-scales; compare the
dashed and continuous lines in Fig.~\ref{fig:auto12} where in the
end-to-end distance the relaxation time is $\sim10^6$ while in the
first-to-second distance is $\sim10^1$. This disparity is simply a
result of the projection that determines the relative contribution of
different eigenfunctions.

\section{MD Simulation details}
$177$ trajectories $1~\mu$s long trajectories of the WW-domain of the human Pin1 Fip (2F21) mutant
were generated using the
GROMACS 4.5 software package~\cite{pronk_gromacs_2013} with the Amber
ff99SB-ILDN force field~\cite{lindorff-larsen_improved_2010} and the
TIP4P-Ew water model~\cite{horn_development_2004}. The starting
structure was taken from the PDB entry
2F21~\cite{jager_structure-function-folding_2006} and considered only
its WW-domain. Energy minimization was performed using steepest
descent for $5\cdot10^4$ steps.  The hydrogen atoms  were  described
by  virtual  sites.  In each trajectory the protein  was positioned
within a triclinic  water box using gmx-solvate, such that the smallest
distance between protein surface and box boundary was larger than
$1.5$~nm. Sodium and chloride ions were added to neutralize the
system, corresponding a physiological concentration of $150$~mmol/l.
The system was first equilibrated for $0.5$~ns in the NVT ensemble,
and subsequently for $1.0$~ns in the NPT ensemble at $1$~atm pressure
and temperature $300$~K, both using an integration time step of
$2$~fs.  The velocity rescaling thermostat~\cite{bussi_canonical_2007}
and Parrinello-Rahman pressure
coupling~\cite{parrinello_polymorphic_1981} were used with coupling
coefficients of $\tau=0.1$~ps and $\tau= 1$~ps, respectively.  All
bond lengths of the solute were constrained using LINCS with an
expansion order of 6, and water geometry was constrained using the
SETTLE algorithm.  Electrostatic interactions were calculated using
PME~\cite{darden_particle_1993}, with a real space cutoff of
$10$~{\AA} and a Fourier spacing of $1.2$~{\AA}. The integration
time-step was $4$~fs, and the coordinates of the alpha carbons were
saved every $10$~ps.  

\section{Fraction of native contacts}
The dynamics of WW-domain of the human Pin1 Fip mutant was projected
on the fraction of native contacts as the reaction coordinate, defined in~\cite{best_native_2013} as
\begin{equation}
    q(t) = \frac{1}{N} \sum_{(i,j) \in S} \frac{1}{1 + \exp[\beta(r_{ij}(t) - \lambda r_{ij}^0)]};
\end{equation}
where $r_{ij}(t)$ is the distance between atoms $i$ and $j$ at time $t$, $r^0_{ij}$ is the same distance in the native state, $S$ is the set of all pairs of the $N$ heavy atoms $(i,j)$ belonging to residues $\theta_i$ and $\theta_j$ such that $|\theta_i - \theta_j| > 3$~{\AA} and $r^0_{i,} < 4.5$~{\AA}. The parameter $\beta=5$~{\AA}$^{-1}$ is a smoothing parameter while $\lambda=1.8$ takes into account the fluctuations of the system. This reaction coordinate was extracted from the files containing the Molecular Dynamics trajectories using the MDTraj library~\cite{mcgibbon_mdtraj_2015}.

\section{Results for long MD simulations}
\begin{figure}
    \centering
    \includegraphics{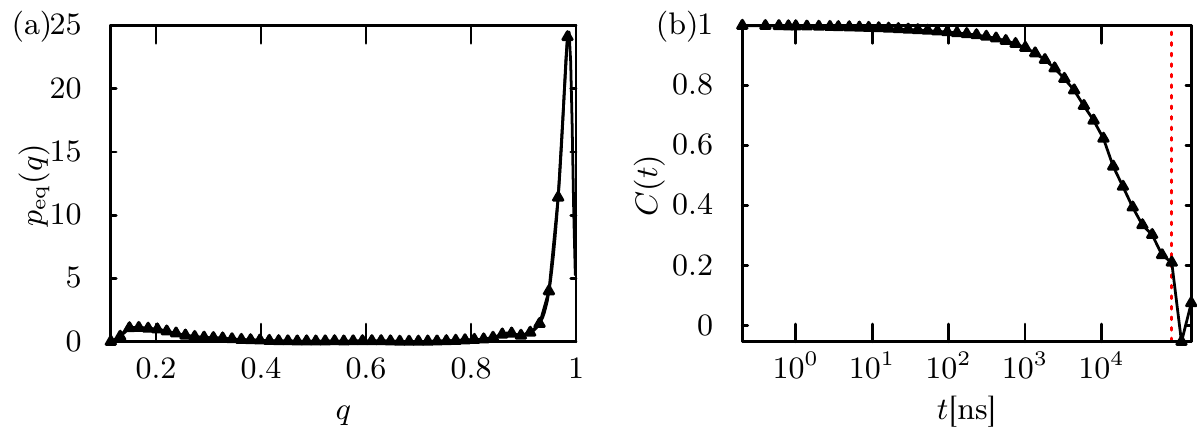}
    \caption{(a) Equilibrium probability density function $p_{\rm eq}(q)$; the peak
      corresponding to the folded configuration is clearly visible and
      comparable with the corresponding peak in the equilibrium
      probability density of the shorter MD simulation of the protein
      in the folded folded-state. The second, smaller peak
      corresponding to the unfolded configurations is noticeable as
      well. (b) Autocorrelation function $C(t)$; at long times
      (e.g.\ beyond the dashed red line)
      insufficient data does not allow for a reliable calculation of $C(t)$.}
    \label{fig:DES_AE}
\end{figure}

The equilibrium probability density $p_{\rm eq}(q)$ and
autocorrelation function $C(t)$ of the fraction of native contacts
determined from the two longer MD trajectories provided by the Shaw
group is shown in Fig.~\ref{fig:DES_AE}. Clearly, $q_t$ does not relax
during the simulation despite the beyond impressive length of the
trajectory. Moreover, because the major change in $q$ is due to the
folding process the intermediate plateau corresponding to the
local equilibrium in the folded state is not visible, as it
contributes negligibly to the total relaxation process. 

\begin{figure}
    \centering
    \includegraphics{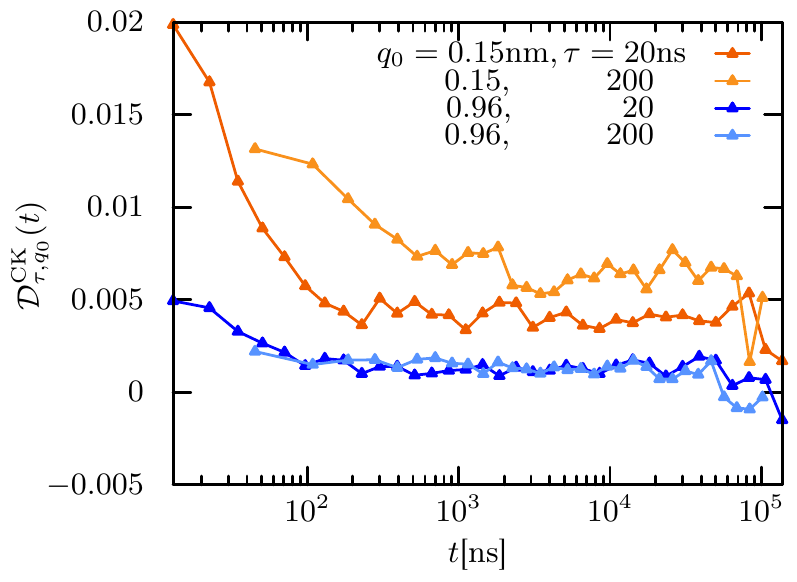}
    \caption{$\mathcal{D}^{\rm CK}_{q_0}(t)$ determined from the Shaw
      simulation data. Memory is found only for lag times $\tau$ of
      $20$~ns. For larger $\tau$ their presence is negligible. For
      $q_0=0.15$ the results are noisier and less reliable due to to
      the poor sampling of the unfolded configurations (see Fig~\ref{fig:DES_AE}a).}
    \label{fig:KLCKDES}
\end{figure}

Despite limited statistics at long times we used the
Chapman-Kolmogorov construction (since this method does not require
that $q_t$ equilibrates) to asses the presence of memory in the
reaction coordinate. The results are depicted in Fig~\ref{fig:KLCKDES}.

Signatures of memory are present only on short time-scales $<100$~ns,
and are the strongest in the deep well corresponding to the folded
state. We therefore confirm that the folding-unfolding transition that
develops on time-scales larger than $1~\mu$s is effectively
memory-less~\cite{best_native_2013} (note that the experimental
unfolding time was estimated to be
$5.7~\mu$s~\cite{piana_computational_2011} while the Molecular
Dynamics simulations yield a value of
$21~\mu$s~\cite{lindorff-larsen_how_2011}). Conversely, both data-sets
show a pronounced memory in the folded-state relaxation.

\section{Estimation of the diffusion coefficient}
We estimate the ($q$-independent)
diffusion coefficient $D(q)$ from a time-series using the first two moments of the
local displacements according to the thermodynamically consistent anti-It\^o convention. We first determine the first and second
moment of the displacement in each bin-point $q_l$ after a single time-step
$\Delta t$ (that is $2.5$ $\mu$s for the hairpin and $10$ ps in the case of the protein), i.e. $\langle \delta q_{\Delta t}^2(l)\rangle$ and
$\langle \delta q_{\Delta t}(l)\rangle$ where $\delta q_{\delta
  t}(l)=q_{t+\Delta t}-q_{t}|_{q_{t+\Delta t}=q_l}$.
\begin{equation}
D(q)=\frac{\langle \delta q_{\Delta
    t}^2(l)\rangle-\langle \delta q_{\Delta t}(l)\rangle^2}{2\Delta t},
\end{equation}
where the brackets $\langle\cdot\rangle$ here denote the average over
all displacements in the bin observed during the entire
time-series.
We consider two bin-sizes, $l_D=$0.01 nm and
$l_D=$0.001 nm, and find the result to be essentially independent on the
precise value of $l_D$ we choose.

\begin{figure}
 \centering
 \includegraphics[width=0.5\textwidth]{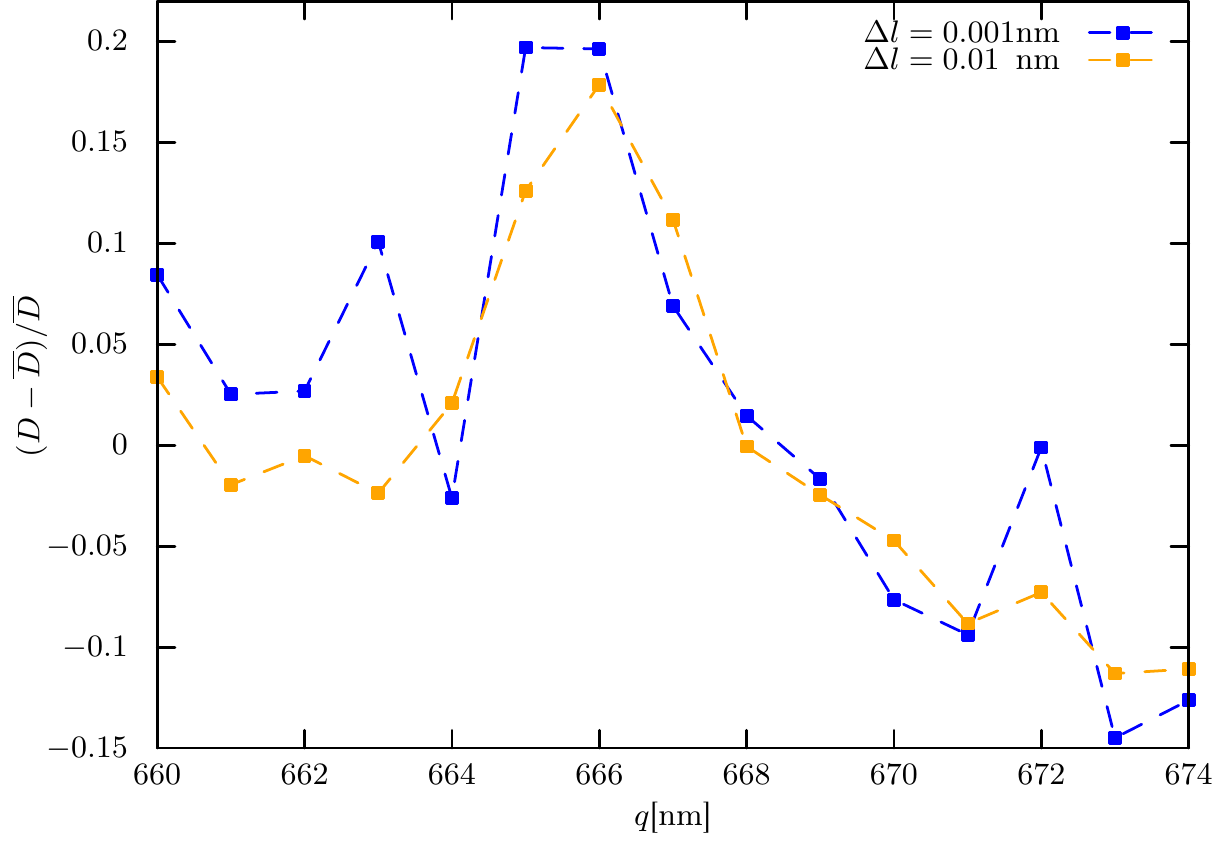}
 \caption{Relative deviation of the local diffusion coefficient in a
   given bin $D_l$ from the average value $\overline{D}=N^{-1}\sum_{l=1}^ND_l$ as a function
   of the position of the bin. In the case of $\Delta l=$ 0.001 nm we
   find $\overline{D}$ = 447 nm$^2/$ms with a deviation $\pm
   9$ nm$^2/$ms and for $\Delta l=$ 0.01 nm we
   find $\overline{D}$ = 448 nm$^2/$ms with a deviation $\pm
   9$ nm$^2/$ms. In a first approximation the values of $D_l$ are
   independent of $l$ and we thus set $D\approx\overline{D}\simeq
   448$ nm$^2/$ms.}
 \label{diffusion}
\end{figure}

In the case of the hairpin the results are rather independent of the
location of the bin $q_l$ (see Fig.~\ref{diffusion}), implying that to a good approximation $D$
may indeed be taken as being constant, such that we instead take
$D(q_l)\to \overline{D}=\sum_{l=1}^{N_b}D(q_l)/N_b$. 

\begin{figure}
    \centering
    \includegraphics{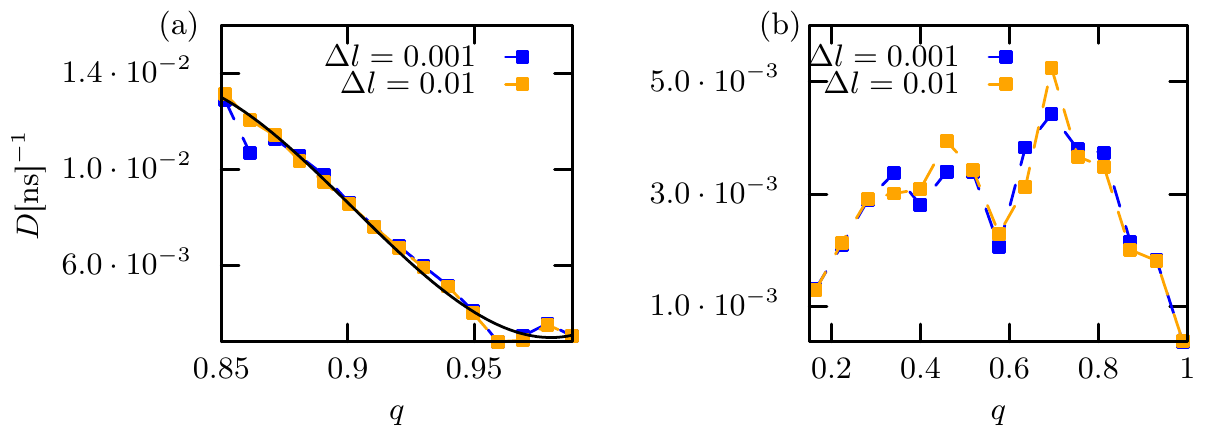}
    \caption{The diffusion landscape $D(q)$ extracted from the shorter
      (left) and longer (right) simulations 
      for two different bin sizes. The black line is the polynomial
      fit (i.e.\ Eq.~\eqref{fit}). }
    \label{fig:diffusion protein}
\end{figure}

In the case of the protein we determine $D(q)$ for both, the shorter
and longer simulation. 
In both cases the diffusion coefficient is found to be 
weakly dependent on $q$, and is smaller in
the folded state, in agreement with the results presented
in~\cite{best_coordinate-dependent_2010}. In order to
efficiently simulate the constructed Markovian process for the shorter simulation
(which attains a local equilibrium), we
fit diffusion landscape to a cubic polynomial
\begin{equation}
D(q)=[-4.03867+13.66777\,q -15.26772\,q^2 + 5.64218\,q^3]\,~{\rm ns}^{-1}.
\label{fit}
\end{equation}
The result are shown in Fig.~\ref{fig:diffusion protein}.

\section{Uncertainty estimation}
We estimated the uncertainty in the computation of the
Kullback-Liebler divergences by considering $M=20$ randomly reduced
the data-sets, each containing $100$ different trajectories
(i.e.\ taking only $\sim 56\%$ of the total number of
trajectories) for the protein, and $40$ different trajectories
(i.e.\ taking only  $\sim 80\%$ of the total number of
trajectories)  for the DNA-hairpin. From these results we determined the standard deviation
in $\mathcal{D}(t)$ as
\begin{equation}
    \sigma_\mathcal{D}(t)=\sqrt{\frac{1}{M}\sum_{i=1}^M (\mathcal{D}(t)-\langle\mathcal{D}(t)\rangle)^2}.
\end{equation}
This analysis was not feasible for the longer MD trajectories provided
by the D.E. Shaw group due to insufficient data. 


\end{document}